\documentclass[12pt]{iopart}
\usepackage{graphicx}

\providecommand{\boldsymbol}[1]{\mbox{\boldmath $#1$}}

\begin{document}
\title{Gravitational radiation from pulsar glitches}

\author{C. A. van Eysden$^{1}$ and A. Melatos$^{1}$}

\address{
$^{1}$School of Physics,
University of Melbourne, Parkville, VIC 3010, Australia.
}

\begin{abstract}
The nonaxisymmetric Ekman flow excited inside a neutron star
following a rotational glitch is calculated analytically
including stratification and compressibility.
For the largest glitches, the gravitational wave strain produced by the hydrodynamic mass quadrupole moment
approaches the sensitivity range of advanced long-baseline
interferometers.
It is shown that the viscosity, compressibility, and orientation 
of the star can be inferred in principle from the width and amplitude ratios
of the Fourier peaks (at the spin frequency and its first harmonic) observed
in the gravitational wave spectrum in the $+$ and $\times$ polarizations.
These transport coefficients constrain the equation of state of
bulk nuclear matter, because they depend sensitively on the degree of
superfluidity.
\end{abstract}


\section{Introduction 
 \label{sec:gligw1}}
Glitches are tiny, impulsive angular accelerations observed in
rotation-powered pulsars.
In total, 287 glitches have been discovered in 101 radio pulsars
over four decades of radio pulse timing experiments,
the majority since the advent of the Parkes Multibeam Survey,
\cite{man01}
improved multifrequency ephemerides,
and better interference rejection algorithms.
\cite{hob02,kra03,jan06}
Glitches also occur occasionally in anomalous X-ray pulsars,
which are believed to be ultramagnetized neutron stars (magnetars).
\cite{dal03,kas03}
The glitch population has been analyzed statistically
from several standpoints.
\cite{mck90,she96,ura99,wan00,hob02,kra03,jan06,per07,mel07}
The fractional increase in angular velocity spans
a wide range, $10^{-11} \leq \delta\Omega/\Omega \leq 10^{-4}$,
suggesting that the angular momentum stored in the superfluid core
of the neutron star is transferred erratically to the solid crust
($\sim 1\%$ of the total moment of inertia)
via a series of discrete coupling events of varying strength. \cite{lyn00,mel07}
While the trigger is unknown, it is often ascribed 
to collective unpinning of quantized superfluid vortices,
e.g. when crust-superfluid differential rotation
(and hence the Magnus force) exceeds a threshold,
\cite{and75}
or when a starquake occurs, heating the crust and boosting the rate of
vortex creep,
\cite{lin96}
although there are dissenting views.
\cite{jon98}
Vortex unpinning is followed by an exponential relaxation phase,
during which crust-superfluid corotation is restored by viscous forces.
\cite{rei92,per05}

It is natural to speculate that glitches excite nonaxisymmetric motions,
generating gravitational radiation.
There are several plausible scenarios in which this can happen.
(1) 
A bulk two-stream instability occurs in the core,
mediated by entrainment between the neutron superfluid and charged species.
\cite{and03}
(2) 
A surface two-stream instability occurs at the interface
between the $^1$S$_0$- and $^3$P$_2$-paired neutron superfluids.
\cite{mas05}
(3) 
The crust cracks, tilts, and precesses.
\cite{jon02,per07}
(4) 
Crust-core differential rotation drives meridional circulation
and, at high Reynolds number, superfluid turbulence.
\cite{per05,per06a,per06b,per07,and07}
(5) 
Vortex unpinning excites
pulsation modes in the multi-component core superfluid,
e.g.\ acoustic, gravity, and Rossby waves,
\cite{rez00,and01}
and quasiradial oscillations.
\cite{sed03,sed06}
On the basis of general arguments, invoking the conservation of energy and angular momentum, the gravitational wave strain is predicted to reach $h \sim 10^{-26}$
from pulsars that glitch strongly and frequently, like Vela.
\cite{and01,sed03}

In this paper, we calculate analytically the gravitational wave signal generated during the relaxation phase of a glitch, when the fluid interior of a neutron star responds to an impulsive, nonaxisymmetric angular acceleration of the crust.  
We idealize the spin-up event as an enduring step increase in the angular velocity of a semi-rigid, fluid-filled container, deliberately simplifying the initial and boundary conditions to keep the focus on how effects like stratification and compressibility (see below) show up in gravitational-wave observables.  (Besides, the detailed microphysics of the spin-up event is unknown in most of the scenarios listed in the previous paragraph.)  
For example, we treat the interior as a single Newtonian fluid with an average viscosity and density, whereas in reality it is a multi-component superfluid with viscous and inviscid components, whose properties vary with depth.  
We therefore neglect the two-stream dynamics in scenarios (1) and (2), nor do we allow for an interface between the inner crust and outer core [cf. scenario (2)].  
The circulation of a rotating superfluid is carried by quantized vortices, which can pin  metastably in between nuclear lattice sites, or at defects (e.g., dislocations, grain boundaries), in the crust.  \cite{don06,avo07}
Yet we neglect the effects of vortex pinning in our model [cf. scenario (5)], except to note that pinning introduces an ambiguity in the boundary conditions, which is not yet resolved in experiments with terrestrial superfluids, let alone in a neutron star. (We expand on this issue in \S\ref{sec:gligw2d} below.)  
Finally, we do not allow the crust to precess [cf. scenario (3)], partly to keep the analysis tractable, and partly because precession of the coupled crust-core system is tied up intimately with the strength of vortex pinning \cite{sha77}, which we neglect.  (In general, the crust possesses a misaligned mass quadrupole moment, whose gravitational wave signal must be added independently to the glitch signal we compute.)  
Regrettably, there is no published literature on the nonlinear evolution of the instabilities which underlie the above glitch scenarios, whether multi-fluid modes in (1)--(4) or vortex Kelvin waves in (5);  the relevant numerical simulations have not yet been performed.  Yet the nonlinear saturation state of the instability determines the degree of nonaxisymmetry (and hence the gravitational wave spectrum) of the spun up container in reality.  \cite{che88, sed99}  This shortcoming is felt most keenly in scenario (4), where numerical simulations are known to reveal nonaxisymmetric meridional circulation even in the small-shear limit \cite{per05,per06a}, an effect excluded artificially by the linear analysis in this paper.

With these approximations in mind, we construct a toy spin-up model in \S\ref{sec:gligw2} by solving the nonaxisymmetric Ekman problem 
using the traditional method of multiple scales,
generalizing previous axisymmetric analyses.
\cite{eas79,abn96}
We next investigate the effects of stratification (\S\ref{sec:gligw3}) 
and compressibility (\S\ref{sec:gligw4}).  
In \S\ref{sec:gligw5}, we compute the gravitational wave strain
generated by the mass quadrupole.
The current quadrupole, which is typically smaller, 
is evaluated in a forthcoming paper.
In \S\ref{sec:gligw6}, \S\ref{sec:gligw8}, 
we show that the amplitude, frequency, phase, quality factor, and polarization can be inverted to measure 
the viscosity and compressibility of the stellar interior
and hence infer its state of superfluidity.
Gravitational radiation from glitches therefore probes directly
the bulk properties of quantum condensates at nuclear densities
$\sim 10^{15}\,{\rm g\,cm^{-3}}$.  The latter topic is attracting growing attention, both theoretically \cite{vre97,gle00,pie04,web07,lat07,kla07} and in heavy-ion collision experiments \cite{har06,for07,ada07,mat07}.

\section{Nonaxisymmetric Ekman flow after a glitch
 \label{sec:gligw2}}
During a glitch, the angular velocity of the stellar crust increases by a small amount
$\delta\Omega \ll \Omega$, on a time-scale that is unresolved by
observations \cite{mcc90} (cf.\ some Crab glitches \cite{won01}).
In reality,
the angular acceleration of the crust is likely to be nonaxisymmetric to some degree,
as argued in Section 1.
Consequently, the resulting spin-up (Ekman) flow is likely to be
nonaxisymmetric as well.
In this section, we calculate the Ekman flow analytically,
generalizing previous axisymmetric analyses. \cite{wal69,abn96}
The exact flow pattern is affected strongly by stratification and compressibility.
These properties, which directly probe the physics of bulk nuclear matter,
therefore leave their imprint on the resulting gravitational wave signal.

The initial and boundary conditions adopted to describe the spin-up event hydrodynamically are discussed critically in \S\ref{sec:gligw2d} and \S\ref{sec:gligw2f}.  We adopt take the standard form favored in the neutron star literature \cite{eas79, abn96, per05, bil00}: a permanent, discontinuous spin-up of a fluid-filled container, which is semi-rigid to accommodate nonaxisymmetry.  These conditions have a long history of being used successfully as the simplest way to capture the spirit of a glitch in a toy model. Nevertheless, they are plainly not self-consistent in three important respects, discussed furthur in \S\ref{sec:gligw2d} and \S\ref{sec:gligw2f}.  (1) The physics of the glitch trigger is unknown, but it is likely to involve vortex pinning, which these hydrodynamic conditions neglect.  (2) The superfluid interior couples frictionally to the solid crust in a way that allows partial slippage at the interface \cite{kha65, rei93, bar00}.  The degree of slippage is still not characterised fully in laboratory experiments with liquid helium, let alone in neutron stars.  (3) The back reaction from vortex turbulence and magnetic fluxoids on the Ekman boundary layer, which contributes to the boundary conditions on the interior flow, remains an unsolved problem \cite{per05,per06b,and07,sed99,men01,alp96,che88}.  In view of these unavoidable uncertainties, the idealizations in the model (e.g. semi-rigidity) are tolerable.

For simplicity, we treat the spin up of a cylinder, following Ref. \cite{abn96}. 
The nonaxisymmetric Ekman problem in a sphere has not yet been solved
analytically in general, although some progress has been made in special limits. \cite{gre68}

\subsection{Model equations 
 \label{sec:gligw2a}} 
Consider a cylinder of height $2L$ and radius $L$,
containing a compressible Newtonian fluid with uniform kinematic viscosity $\nu$.
(Note that the effective (eddy averaged) $\nu$ may be boosted substantially by turbulent Reynolds stresses, as in terrestrial oceanic or atmospheric flows \cite{mel07b, ped83}).
Initially, the cylinder rotates
with angular velocity $\vec{\Omega}=\Omega\vec{e}_z$ about the $z$ axis.
In the rotating frame, the compressible Navier-Stokes equation,
including Coriolis and centrifugal forces, takes the form
\begin{equation}
 \frac{\partial \vec{v}}{\partial t} + \vec{v} \cdot \nabla \vec{v} +
  2\vec{\Omega}\times\vec{v}
 = 
 - \frac{1}{\rho} \nabla p + \vec{g}+ \nu\nabla^2 \vec{v}+\frac{\nu}{3}\nabla\left( \nabla\cdot\vec{v}\right) 
  + \nabla\left(\frac{1}{2}\Omega^2r^2\right)~.
\label{eq:gligw1}
\end{equation}
The symbols $\vec{v}$, $\rho$, $p$, and $\vec{g}$
denote the fluid velocity, fluid density, 
pressure, and gravitational acceleration respectively.
Magnetic forces are neglected for the sake of simplicity, 
even though they are known to modify the spin up of the core 
(which is coupled to the crust by cyclotron-vortex waves).
\cite{eas79,men98}
We adopt cylindrical coordinates $(r,\phi,z)$ in the rotating frame.
The gravitational acceleration is taken to be uniform and perpendicular
to the midplane of the cylinder, viz.
\begin{equation}
 \vec{g}
 = 
 \left\{ \begin{array}{ll}
  -g \vec{e}_z & \textrm{if $z>0$}\\
  +g \vec{e}_z & \textrm{if $z<0$}~,
\end{array} \right.
\label{eq:gligw2}
\end{equation}
where $g$ is constant.  This form of the gravitational field is standard in both the neutron star literature \cite{eas79,abn96} and the classic papers of geomechanics \cite{ped67,wal69,gre68}.  It is sourced by a singular surface mass distribution at $z=0$ and is therefore nominally unphysical.  However, it leads to a self-consistent spin-up model, renders the problem analytically tractable, and has been validated by comparisons with numerical simulations involving smooth mass distributions \cite{mou07}.
Henceforth, we restrict attention to the region $z\geq 0$, as the flow is symmetric about the midplane of the cylinder.

The fluid satisfies the continuity equation
\begin{equation}
 \frac{\partial \rho}{\partial t} + \nabla \cdot (\rho \vec{v}) = 0~.
 \label{eq:gligw3}
\end{equation}
The energy equation is written in a form that relates the convective derivatives of the pressure and density, viz.
\begin{equation}
 \left( \frac{\partial }{\partial t} + \vec{v}\cdot\nabla \right)\rho  = \frac{1}{c^{2}}\left( \frac{\partial }{\partial t} + \vec{v}\cdot\nabla  \right) p~,
 \label{eq:gligw4}
\end{equation}
where $c^{2}$ is the propagation speed of an acoustic disturbance in the Lagrangian frame, determined by the equation of state (e.g., adiabatic or isothermal).
 
It is useful to recast (\ref{eq:gligw2})--(\ref{eq:gligw4}) in dimensionless form,
when preparing to separate the slow and fast dynamics of the flow.
The characteristic time-scale for Ekman pumping is 
$t_{\rm E}=E^{-1/2}\Omega^{-1}$,
where we define the dimensionless Ekman number 
\begin{equation}
 E=\frac{\nu}{L^2 \Omega }~.
 \label{eq:gligw5}
\end{equation}
The characteristic length-scale is $L$.  The velocity scale for the spin-up flow is $L \delta \Omega $.  The remaining scales are chosen so that the hydrostatic forces dominate inertial forces in the momentum equation.  The pressure scale is taken to be $ p_c=\rho_0 g L$, the density scale $\rho_0$ is taken to be the equilibrium density in the midplane $z=0$.  These choices lead to the following scaled equations:
\begin{eqnarray}
 \epsilon F & \left( E^{1/2} \frac{\partial \vec{v}}{\partial t} + \epsilon \vec{v} \cdot \nabla \vec{v} +
  2\vec{e}_z\times\vec{v} \right) \nonumber \\
 & = 
 - \frac{1}{\rho} \nabla p - \vec{e}_z + \epsilon F E \left[ \nabla^2 \vec{v}+ \frac{1}{3}\nabla\left( \nabla\cdot\vec{v}\right)\right] 
  + F \nabla\left(\frac{1}{2}r^2\right)~,
\label{eq:gligw6}
\end{eqnarray}
\begin{equation}
 E^{1/2} \frac{\partial \rho}{\partial t} + \epsilon \nabla \cdot (\rho \vec{v}) = 0~,
 \label{eq:gligw7}
\end{equation}
\begin{equation}
 \left( E^{1/2} \frac{\partial }{\partial t} + \epsilon \vec{v}\cdot\nabla \right) \rho = K \left( E^{1/2} \frac{\partial }{\partial t} + \epsilon \vec{v}\cdot\nabla  \right)p ~.
 \label{eq:gligw8}
\end{equation}
Equations (\ref{eq:gligw6})--(\ref{eq:gligw8}) introduce three dimensionless quantities: the Rossby number $\epsilon=\delta\Omega / \Omega$, the Froude Number $F=L\Omega^2/g$, and the scaled compressibility $K=gL/c^2$. 

At time $t=0$, the cylinder is accelerated impulsively. If the Rossby number is small, 
we can solve for the equilibrium and spin-up (Ekman) flows separately by making the  
perturbation expansion $\rho \mapsto \rho + \epsilon \delta\rho$, $p \mapsto p + \epsilon \delta p$.  We also replace $\vec{v}$ with $\delta\vec{v}$ since it is of order $\epsilon$.  In the rotating frame, the equilibrium flow is steady and axisymmetric, with $\rho=\rho(r,z)$, $p=p(r,z)$ and
\begin{equation}
 0 = - \frac{1}{\rho} \nabla p - \vec{e}_z + F \nabla\left(\frac{1}{2}r^2\right) ~.
 \label{eq:gligw9}
\end{equation}
On the other hand, the spin-up flow is unsteady and nonaxisymmetric,
with $\delta\rho=\delta\rho(r,\phi,z,t)$, 
$\delta p=\delta p(r,\phi,z,t)$, 
$\delta \vec{v}=\delta\vec{v}(r,\phi,z,t)$, 
\begin{eqnarray}
  F &\left( E^{1/2} \frac{\partial\delta\vec{v}}{\partial t}+ 2 \vec{e}_z \times \delta\vec{v} \right)
  \nonumber \\ & =  - \frac{1}{\rho} \nabla \delta p  - \frac{\delta\rho}{\rho}\vec{e}_z 
+ F E \left[ \nabla^2\delta\vec{v}+\frac{1}{3}\nabla\left( \nabla\cdot\delta\vec{v}\right)\right] 
+ F \frac{\delta\rho}{\rho} \nabla\left(\frac{1}{2}r^2\right)~,
 \label{eq:gligw10}
\end{eqnarray}
\begin{equation}
 E^{1/2} \frac{ \partial \delta \rho}{\partial t} + \nabla \cdot (\rho \delta\vec{v}) = 0~,
 \label{eq:gligw11}
\end{equation}
and
\begin{equation}
 E^{1/2}\frac{\partial \delta \rho}{\partial t} + \delta\vec{v} \cdot \nabla \rho
 = K \left(E^{1/2} \frac{\partial \delta p}{\partial t} + \delta\vec{v} \cdot \nabla p \right) ~.
 \label{eq:gligw12}
\end{equation}
The effects of stratification and compressibility enter the spin-up flow through (\ref{eq:gligw12}), which is sometimes written alternatively as \cite{and01b}
\begin{equation}
 E^{1/2}\frac{\partial \delta \rho}{\partial t} 
 = K E^{1/2} \frac{\partial \delta p}{\partial t} - \rho \delta\vec{v} \cdot \vec{A} ~.
 \label{eq:gligw13}
\end{equation}
The Schwarzchild discriminant, 
\begin{equation}
 \vec{A}= \frac{1}{\rho} \nabla \rho + K \vec{e}_z - K F \nabla\left(\frac{1}{2}r^2 \right) ~,
 \label{eq:gligw14}
\end{equation}
characterises the buoyancy.  The buoyant restoring force experienced by a fluid element when displaced from equilibrium in the $z$ direction is proportional to 
$ -g \partial(\ln{ \rho})/\partial z - g^2 /c^{2} $.
It increases as the density gradient steepens and decreases as the compressibility increases.  The frequency of oscillation of the fluid element is called the Brunt-V\"{a}is\"{a}l\"{a} frequency.  
When scaled to $\Omega$, the Brunt-V\"{a}is\"{a}l\"{a} frequency is related to the Schwarzchild discriminant by $N=\left(-A_z/F \right)^{1/2}$.

\subsection{Equilibrium flow 
 \label{sec:gligw2b}}
To solve (\ref{eq:gligw9}) and (\ref{eq:gligw10})--(\ref{eq:gligw12}) analytically, we follow Refs \cite{abn96} and \cite{wal69} and neglect centrifugal terms, as the Froude number is typically small in a neutron star.  However, we retain all other terms containing $F$ until the end of the calculation, as $F N^2$ and $K$ can be of similar magnitude.  We also assume that $\rho^{-1} d\rho/d z $ is uniform throughout the star to simplify the problem analytically.  The stratification of the equilibrium flow can arise from either compositional or entropic gradients. 

Solving (\ref{eq:gligw9}) we obtain
\begin{equation}
 \rho(z)=e^{-K_s z}
 \label{eq:gligw15}
\end{equation}
\begin{equation}
 p(z)=K_s^{-1} e^{-K_s z}
 \label{eq:gligw16}
\end{equation}
where the dimensionless quantity, $K_s=L/z_s=-L \rho^{-1} d\rho/d z $ is proportional to the reciprocal of the stratification length scale, $z_s$.  The density at the midplane $(z=0)$ is chosen to be the fudicial density scale $\rho_0$, as discussed in \S\ref{sec:gligw2a}.  The pressure at the surface $(z=L)$ can be made to vanish by adding a constant to (\ref{eq:gligw16}), but we leave it nonzero here (without loss of generality) in order to maintain consistency with the assumption of uniform sound speed, e.g. if the flow is adiabatic, we have $c^2=\partial p/\partial \rho=\gamma p/\rho=\gamma g z_s$, where $\gamma$ is the ratio of specific heats.

\subsection{Spin-up flow 
 \label{sec:gligw2c}}
Neglecting the centrifugal force terms, the vector components of (\ref{eq:gligw10})--(\ref{eq:gligw12}) are
\begin{eqnarray}
  F \left( E^{1/2} \frac{\partial \delta v_r}{\partial t}-  2 \delta v_\phi \right)
  = & - \frac{\partial }{\partial r} \left( \frac{ \delta p}{\rho}\right) + F E \nonumber \\
& \times \left[ \left( \nabla^2-\frac{1}{r^2}\right) \delta v_r -\frac{2}{r^2}\frac{\partial \delta v_\phi}{\partial \phi} + \frac{1}{3}\frac{\partial }{\partial r}\left( \nabla\cdot\delta\vec{v}\right)\right] 
~,
 \label{eq:gligw17}
\end{eqnarray}
\begin{eqnarray}
  F \left( E^{1/2} \frac{\partial \delta v_\phi}{\partial t}+  2 \delta v_r \right)
  =&  - \frac{1}{r}\frac{\partial }{\partial \phi} \left( \frac{ \delta p}{\rho}\right) + F E \nonumber \\
& \times \left[ \left( \nabla^2-\frac{1}{r^2}\right) \delta v_\phi +\frac{2}{r^2}\frac{\partial \delta v_r}{\partial \phi}+ \frac{1}{3 r}\frac{\partial }{\partial \phi}\left( \nabla\cdot\delta\vec{v}\right)\right] 
~,
 \label{eq:gligw18}
\end{eqnarray}
\begin{eqnarray}
  F  E^{1/2} \frac{\partial \delta v_z}{\partial t}
  =&  - \frac{\partial }{\partial z} \left( \frac{ \delta p}{\rho}\right) +\left(K_s \frac{ \delta p}{\rho}-\frac{ \delta \rho}{\rho}\right)+ F E \nonumber \\ 
& \times \left[ \nabla^2 \delta v_z + \frac{1}{3}\frac{\partial }{\partial z}\left( \nabla\cdot\delta\vec{v}\right)\right] 
~,
 \label{eq:gligw19}
\end{eqnarray}
\begin{equation}
 E^{1/2} \frac{ \partial }{\partial t} \left(\frac{\delta \rho}{\rho}\right) +  \nabla\cdot\delta\vec{v}  = K_s \delta v_z~,
 \label{eq:gligw20}
\end{equation}
and
\begin{equation}
 E^{1/2}\frac{ \partial }{\partial t} \left(\frac{\delta \rho}{\rho}\right) 
 = K E^{1/2} \frac{ \partial }{\partial t} \left(\frac{\delta p}{\rho}\right) + F N^2 \delta v_z ~,
 \label{eq:gligw21}
\end{equation}
where $N^2=\left(K_s-K\right)/F$.  We now proceed according to the method of multiple scales by expanding $\delta\vec{v}$, $\delta p$ and $\delta \rho $ in perturbation series in the small parameter $E^{1/2}$.
\cite{gre63,wal69,ped67,abn96}
To order $O(E^0)$, we find
\begin{eqnarray}
  \delta v_r^0 
  &=& 
  -\frac{1}{2 F r}\frac{\partial}{\partial \phi}\left(\frac{\delta p^0}{\rho}\right) ~,
  \label{eq:gligw22} \\
  \delta v_\phi^0 
  &=& 
  \frac{1}{2 F}\frac{\partial}{\partial r}\left(\frac{\delta p^0}{\rho}\right) ~,
  \label{eq:gligw23} \\
  \delta v_z^0 
   &=& 
   0~,
  \label{eq:gligw24}
\end{eqnarray}
\begin{equation}
 \nabla\cdot\delta\vec{v}^0  = K_s \delta v_z^0~,
 \label{eq:gligw25}
\end{equation}
\begin{equation}
 \delta\rho^0=-\frac{\partial \delta p^0}{\partial z}~,
 \label{eq:gligw26}
\end{equation}
where the superscript `0' denotes a zeroth-order quantity.
In view of (\ref{eq:gligw22})--(\ref{eq:gligw24}), 
the $O(E^0)$ continuity equation (\ref{eq:gligw25}) is identically satisfied. 
To order $O(E^{1/2})$, we find
\begin{eqnarray}
  \delta v_r^1 
  &= & 
  \frac{1}{4 F}\frac{\partial \Phi}{\partial r}
  -\frac{1}{2 F r}\frac{\partial}{\partial \phi}\left(\frac{\delta p^1}{\rho}\right) ~,
  \label{eq:gligw27} \\
  \delta v_\phi^1 
  &= & 
  \frac{1}{4 F r}\frac{\partial \Phi}{\partial \phi}
  +\frac{1}{2 F}\frac{\partial}{\partial r}\left(\frac{\delta p^1}{\rho}\right) ~,
  \label{eq:gligw28} \\
  \delta v_z^1 
  &= & 
   \frac{1}{F N^2}\frac{\partial \Phi}{\partial z} - \Phi~,
  \label{eq:gligw29}
\end{eqnarray}
\begin{equation}
 \frac{\partial}{\partial t} \left( \frac{\delta \rho^1}{\rho} \right) +\nabla\cdot\delta\vec{v}^1  = K_s \delta v_z^0~,
 \label{eq:gligw30}
\end{equation}
where we define $\Phi = - \partial \left( \delta p^0/\rho\right)/ \partial t$,
and the superscript `1' denotes a first-order quantity.
Upon substituting (\ref{eq:gligw22})--(\ref{eq:gligw29}) into the
$O(E^{1/2})$ continuity equation (\ref{eq:gligw30})
we arrive at
\begin{equation}
  \frac{1}{r}\frac{\partial}{\partial r}\left( r \frac{\partial \Phi}{\partial r} \right)
  + \frac{1}{r^2}\frac{\partial^2 \Phi}{\partial \phi^2}
  -\frac{4 K_{s} }{N^2}\frac{\partial \Phi}{\partial z}
 +\frac{4}{N^2}\frac{\partial^2 \Phi}{\partial z^2}=0~.
 \label{eq:gligw31}
\end{equation}
Equation (\ref{eq:gligw31}) is solvable by separation of variables.  It is slightly more general than the equation for $\Phi$ in Ref. \cite{abn96} as it allows for the possibility that $F N^2$ and $K$ are of similar magnitude, which is likely in a neutron star.

\subsection{Boundary conditions
 \label{sec:gligw2d}}
The boundary conditions on $\Phi$ are determined by the boundary conditions on the velocity fields, given by equations (\ref{eq:gligw22}) and (\ref{eq:gligw23}).  Note that the boundary conditions are imposed on the $O(E^0)$ flow.  To impose boundary conditions on the $O(E^{1/2})$ flow, one must specify $\delta p^1$ and utilise (\ref{eq:gligw27}) and (\ref{eq:gligw28}).  The general solution to (\ref{eq:gligw31}) has the form
\begin{equation}
 \Phi(r,\phi,z,t) 
 = F \sum^{\infty}_{m=0} \sum^{\infty}_{n=1} J_m(\lambda_{mn} r) \left[ A_{mn}(t)\cos(m\phi)+B_{mn}(t)\sin(m\phi)\right] Z_{mn}(z)~,
 \label{eq:gligw32}
\end{equation}
where $m\geq 0$ is an integer, $\lambda_{mn}$ is the $n$-th root of $J_m(\lambda)=0$, and $A_{mn}$ and $B_{mn}$ are  coefficients which are functions of time. The prefactor $F$ is extracted explicitly in anticipation that $\Phi$ is a quantity of order $F$.  The solution is periodic in $\phi$, regular in the limit $r\rightarrow 0$,
and vanishes at the cylinder wall $r=1$.  

To specify the boundary conditions on the axial component of the velocity, we require that the flow be symmetric about the midplane of the cylinder, viz. $ \delta v_z^1=0$ at $z=0$.  The final boundary condition on $Z_{mn}$ is arbitrary, and we choose $Z_{mn}(1)=1$.  Hence we find
\begin{equation}
 Z_{mn}(z)=\frac{\left(F N^2 - \beta_{-} \right) e^{\beta_{+}z} - \left(F N^2-\beta_{+}\right) e^{\beta_{-}z}}{\left(F N^2-\beta_{-}\right)e^{\beta_{+}}-\left(F N^2-\beta_{+}\right)e^{\beta_{-}}}~,
 \label{eq:gligw33}
\end{equation}
with
\begin{equation}
 \beta_\pm =\frac{1}{2}\left[ K_{s} \pm \left( K_{s}^2+N^2 \lambda^2_{mn}\right)^{1/2}\right]~.
 \label{eq:gligw34}
\end{equation}

The boundary conditions for the axisymmetric mode $(m=0)$ at $r=1$ differ from those of previous authors \cite{wal69,abn96}.  For the axisymmetric mode, the zeroth-order radial flow is zero everywhere by virtue of (\ref{eq:gligw22}), and Refs. \cite{wal69,abn96} impose no penetration on the first-order radial flow (\ref{eq:gligw27}), giving $\partial \Phi /\partial r =0$ at $r=1$.  However, for the higher order nonaxisymmetric modes $\left(m>0\right)$, we are forced, by (\ref{eq:gligw22}), to demand $\partial \Phi /\partial \phi =0$ at $r=1$, so that the no-penetration boundary condition on the zeroth order radial flow is satisfied.  For generality, we impose the same boundary condition on the axisymmetric mode as on the nonaxisymmetric modes, namely $\partial \Phi /\partial \phi =0$ at $r=1$.  The resultant solution qualitatively resembles that found in Refs. \cite{wal69,abn96}.

The above boundary conditions, based on a single Newtonian fluid with an average viscosity, are grossly idealized.  In fact, the liquid interior of the star is a multi-component mixture of charged, viscous fluid and uncharged, inviscid superfluid. The boundary conditions satisfied by the superfluid mixture are uncertain, not just in neutron stars, but also in terrestrial experiments with liquid helium.  A large body of numerical and experimental research \cite{kha65, hil77, bar00} suggests that the behaviour of the superfluid lies somewhere between no slip and perfect slip where it touches the container, with the exact amount of slip determined by the density of the (turbulent) quantized vortices in the boundary layer, which is poorly known.  In a neutron star, this uncertainty is magnified by the presence of superconducting magnetized fluxoids, which modify the structure of the boundary layer in a nonaxisymmetric way.  \cite{sed99, men01, che88}  Moreover, the velocity components at the container are partly determined by pinning, which this paper neglects, but which is predicted to vary in strength across macroscopic capacitive domains \cite{che88, war08}.

\subsection{Temporal evolution
 \label{sec:gligw2e}}

In Ekman pumping,
the temporal evolution of the interior flow is controlled by the 
boundary conditions at the top and bottom faces of the cylinder.
The mass flux into (out of) the thin viscous boundary layer is related to
the circulation just outside the layer by
\cite{wal69,ped67,abn96}
\begin{eqnarray}
 \left. \delta v_z \right|_{z=\pm 1}= \left. \mp \frac{1}{2} E^{1/2} ( {\nabla}\times \delta\vec{v} )_z \right|_{z=\pm 1}~.
 \label{eq:gligw35}
\end{eqnarray}
This relation follows from the structure of the Ekman layer in an incompressible flow, but it applies equally when the flow is compressible; the fourth term on the right hand side of (\ref{eq:gligw1}) introduces corrections of $O(E)$ to (\ref{eq:gligw35}).  Upon differentiating (\ref{eq:gligw35}) with respect to time and substituting
(\ref{eq:gligw22}), (\ref{eq:gligw23}), and (\ref{eq:gligw32}), 
we obtain
\begin{equation}
 \left. \frac{\partial \delta v_z^1}{\partial t}\right|_{z= \pm 1}=\mp \frac{1}{4}\lambda_{mn}^2\Phi(r,\phi,\pm1,t)~.
 \label{eq:gligw36}
\end{equation}
Combining (\ref{eq:gligw29}) and (\ref{eq:gligw36}), we find that the $(m,n)$-th Ekman mode relaxes exponentially on the
Ekman time-scale as
$\Phi \propto \exp(-\omega_{mn} t)$, with
\cite{abn96}
\begin{equation}
 \omega_{mn}
 =\frac{\lambda_{mn}^2\left[\left(F N^2 -\beta_{-}\right)e^{\beta_{+}}-\left(F N^2-\beta_{+}\right)e^{\beta_{-}}\right]}{\left(4 F K+\lambda_{mn}^2\right)\left(e^{\beta_{+}}-e^{\beta_{-}}\right)}
 \label{eq:gligw37}
\end{equation}
The general solution for the pressure perturbation 
can now be written down:
\begin{eqnarray}
 \frac{\delta  p^0 (r,\phi,z,t)}{\rho(z)}&=& \frac{\delta p^0_\infty (r,\phi,z)}{\rho(z)}\nonumber \\ 
 & &
-F \sum_{m=0}^\infty \sum_{n=1}^\infty
 J_m(\lambda_{mn}r) 
 \left( A_{mn}\cos{m\phi}+B_{mn}\sin{m\phi}\right) \nonumber \\
& & \times Z_{mn}(z) e^{-\omega_{mn} t}~.
 \label{eq:gligw38}
\end{eqnarray}
In (\ref{eq:gligw38}), the function $\delta p_\infty^0(r,\phi,z)$ corresponds to the steady-state pressure profile of the spun up cylinder. All that remains is to determine the coefficients $A_{mn}$ and $B_{mn}$, which are now constants, by specifying the initial and final conditions of the spin up.

\subsection{Initial conditions
 \label{sec:gligw2f}}
If the pressure perturbation is zero initially,
then $A_{mn}$ and $B_{mn}$ are determined by evaluating (\ref{eq:gligw38})
at $z=1$.
Specifying the steady-state solution $\delta p^0_\infty/\rho$ is equivalent to specifying the steady-state velocity fields obeying (\ref{eq:gligw22}) and (\ref{eq:gligw23}).  We therefore require a steady-state, nonaxisymmetric solution to the Navier-Stokes equations, namely
\begin{equation}
 \frac{\delta p^0_\infty(r,\phi,1)}{\rho(1)}
 =F \sum_{m=0}^\infty
  r^m (r^2-1) \cos(m\phi)~,
 \label{eq:gligw39}
\end{equation}
which ensures that there is no penetration at $r=1$,
and that the flow vanishes smoothly on the rotation axis.  For $m=0$, the familiar steady-state, axisymmetric flow is recovered \cite{abn96,wal69}.
The corresponding velocity boundary conditions on the top and bottom faces of the cylinder are (in dimensional form)
\begin{equation}
 \delta v_r(r,\phi,L,t)=\frac{m \delta \Omega L}{2}\left(\frac{r}{L}\right)^{m-1}\left[\left(\frac{r}{L}\right)^2-1\right]\sin{m \phi}~,
 \label{eq:gligw39a}
\end{equation}
\begin{equation}
 \delta v_\phi(r,\phi,L,t)=\frac{ \delta \Omega L}{2}\left(\frac{r}{L}\right)^{m-1}\left[\left(m+2\right)\left(\frac{r}{L}\right)^2-m\right]\cos{m \phi}
 \label{eq:gligw39b}~.
\end{equation}
The boundary conditions on the side walls are given by (\ref{eq:gligw39a}) and (\ref{eq:gligw39b}) evaluated at $r=L$.  The above choice is somewhat artificial, because it implies that the 
top and bottom faces of the cylinder do not rotate rigidly,
as one might expect for the crust of a neutron star.  
However, this shortcoming is not serious when compared with the other uncertainties identified at the start of \S\ref{sec:gligw2} and in \S\ref{sec:gligw2d} (e.g. the degree of boundary layer slippage, and inhomogeneous pinning \cite{che88}). Indeed, the semi-rigid flow boundary conditions described by (\ref{eq:gligw39a}) and (\ref{eq:gligw39b}) is a reasonable model for the outer edge of the interior flow which is isolated from the rigid container by the turbulent Ekman boundary layer. The model captures qualitatively the main elements of the astrophysical problem: a step increase in angular velocity, which in turn excites a nonaxisymmetric internal flow.

The Fourier coefficients $A_{mn}$ in (\ref{eq:gligw41})--(\ref{eq:gligw45}) can be computed from
\begin{equation}
 A_{mn}
 =
 \frac{2}{J_{m+1}^2(\lambda_{mn})}\int_{0}^{1}dr \,
   r^{m+1} (r^2-1) J_m(\lambda_{mn}r)~;
 \label{eq:gligw40}
\end{equation}
the first few are 
$A_{11}=-0.706$, $A_{12}=0.154$, $A_{21}=-0.521$, and $A_{22}=0.148$.  Given (\ref{eq:gligw39}), we have $B_{mn}=0$.

The steady-state solution (\ref{eq:gligw39})--(\ref{eq:gligw39b}) implies that the cylinder spins up to a nonaxisymmetric steady state (which continually emits gravitational radiation; see \S\ref{sec:gligw5}). In reality, it is indeed likely that some nonaxisymmetry survives between glitches, as a result of inhomogeneous vortex pinning in macroscopic domains. \cite{alp96, won01, che88}  As an alternative, however, one can consider a scenario in which nonaxisymmetric modes are excited by the glitch trigger at $t=0$ (as discussed in \S\ref{sec:gligw1}) and subsequently dissipate as $t\rightarrow \infty$, leaving only the axisymmetric mode.  This scenario can be treated mathematically by specifying the conditions corresponding to (\ref{eq:gligw38}) at $t=0$ and $t\rightarrow\infty$.  It is deferred to future work.

\subsection{Density and velocity fields
 \label{sec:gligw2g}}
With the constants of the Fourier expansion determined, the density, pressure and velocity can be written out explicitly using (\ref{eq:gligw22})--(\ref{eq:gligw30}).  Transforming out of the rotating frame into the observer's frame, and restoring dimensional variables, the complete expressions read

\begin{eqnarray}
 v_r(r,\phi,z,t)
 & = &
 \frac{L^2 \delta\Omega}{2 r}
 \sum_{m=0}^\infty \sum_{n=1}^\infty m A_{mn} J_m(\lambda_{mn}r/L) \sin{m(\phi-\Omega t)}
 \nonumber \\
 & & 
 \times \left[ \frac{\left(F N^2-\beta_- \right) e^{\beta_+ z/L}-\left( F N^2-\beta_+\right) e^{\beta_- z/L}}{\left(F N^2-\beta_-\right) e^{\beta_+}-\left(F N^2-\beta_+\right)e^{\beta_-}} \right] \nonumber \\
 & & \times \left(1-e^{-E^{1/2}\omega_{mn} t} \right)~,
 \label{eq:gligw41}
\end{eqnarray}
\begin{eqnarray}
 v_\phi(r,\phi,z,t)
 & = &
 \Omega r + \frac{L^2 \delta\Omega}{2} \sum_{m=0}^\infty \sum_{n=1}^\infty A_{mn} 
 \frac{\partial}{\partial r}\left[J_m(\lambda_{mn}r/L)\right] \cos{m(\phi-\Omega t)}
 \nonumber \\
 & & 
 \times \left[ \frac{\left(F N^2-\beta_-\right) e^{\beta_+ z/L}-\left(F N^2-\beta_+\right) e^{\beta_- z/L}}{\left(F N^2-\beta_-\right) e^{\beta_+}-\left(F N^2-\beta_+\right)e^{\beta_-}} \right] \nonumber \\
 & & \times \left(1-e^{-E^{1/2}\omega_{mn} t} \right)~,
 \label{eq:gligw42}
\end{eqnarray}
\begin{eqnarray}
 v_z(r,\phi,z,t)
 & = &
 -\frac{L \delta\Omega E^{1/2}}{4} \sum_{m=0}^\infty \sum_{n=1}^\infty A_{mn} \lambda_{mn}^2
 J_m(\lambda_{mn}{r}/{L})
 \nonumber \\
 & & 
 \times \cos{m(\phi-\Omega t)} \left( \frac{e^{\beta_+ z/L}-e^{\beta_- z/L}}{e^{\beta_+}-e^{\beta_-}} \right)
 e^{-E^{1/2}\omega_{mn} t}~,
 \label{eq:gligw43}
\end{eqnarray}
\begin{eqnarray}
 \rho(r,\phi,z,t)
 & = &
 \rho_0 e^{-z/z_s} + \rho_{\rm 0} \frac{\Omega \delta\Omega L}{g} \sum_{m=0}^\infty \sum_{n=1}^\infty A_{mn} J_m(\lambda_{mn}{r}/{L}) \cos[m(\phi-\Omega t)]
 \nonumber \\
 & & 
 \times  \left[ \frac{\left(F N^2-\beta_- \right)\beta_- e^{-\beta_- z/L}-\left(F N^2-\beta_+ \right)\beta_+ e^{-\beta_+ z/L}}{\left(F N^2-\beta_-\right) e^{\beta_+}-\left(F N^2-\beta_+\right)e^{\beta_-}} \right] \nonumber \\
 & & \times \left( 1-e^{-E^{1/2}\omega_{mn} t}\right)~,
 \label{eq:gligw44}
\end{eqnarray}
\begin{eqnarray}
 p(r,\phi,z,t)
 & = &
  \rho_0 g z_s e^{-z/z_s} + \rho_{\rm 0} L^2 \Omega \delta\Omega \sum_{m=0}^\infty \sum_{n=1}^\infty A_{mn}
 J_m(\lambda_{mn}r/L) \cos{m(\phi-\Omega t)}
 \nonumber \\
 & & 
 \times  \left[ \frac{\left(F N^2-\beta_- \right)e^{-\beta_- z/L}-\left(F N^2-\beta_+ \right) e^{-\beta_+ z/L}}{\left(F N^2-\beta_- \right)e^{\beta_+}-\left(F N^2-\beta_+\right) e^{\beta_-}} \right]  \nonumber \\
& & \times \left(1-e^{-E^{1/2}\omega_{mn} t} \right)~.
 \label{eq:gligw45}
\end{eqnarray}

\section{Stratification
 \label{sec:gligw3}}

Stratification acts to restrict the penetration depth of the secondary $O(E^{1/2})$ flow, preventing the interior of the container from spinning up completely.  Because the spun up volume decreases, the Ekman time decreases.  When a fluid element is displaced vertically from its equilibrium position in the hydrostatic density gradient $d\rho/d z$, it experiences a buoyancy force, proportional to $N^2$, as discussed in \S\ref{sec:gligw2a}.  The strength of the buoyancy force is determined in part by the fluid element's ability to adjust its density to match the surroundings, which is related to its compressibility (see \S\ref{sec:gligw4} below).  However, for a fixed compressibility, the $z$-dependent factors in (\ref{eq:gligw41}) and (\ref{eq:gligw42}) indicate that the rotating fluid on the midplane lags behind the final velocity at $z = \pm 1$ as $t\rightarrow \infty$.  The lag increases as $d \rho/d z$ increases.  The interior of the cylinder eventually spins up when vorticity diffuses out of the boundary layer on the much longer time-scale $E\Omega^{-1}$.

To illustrate the physics of stratification, consider the special case of an incompressible Ekman flow.  To investigate this limit, we take $K\rightarrow0$ and hence $N^2\rightarrow K_s/F$. As the Froude number is small we have $K_s \ll N^2$, $\beta_\pm\approx\pm \lambda_{mn}N/2$, and hence

\begin{equation}
 \frac{\omega_{mn}}{\Omega}
 =\frac{\lambda_{mn} N}{2} \coth {\frac{\lambda_{mn} N}{2}}~.
 \label{eq:gligw46}
\end{equation}
Equation (\ref{eq:gligw46}) shows that, in the incompressible limit, an increase in the density gradient reduces the spin up time.  This effect, familiar from axisymmetric flows \cite{wal69}, is enhanced for higher mode numbers $m$.

\begin{figure}[h]
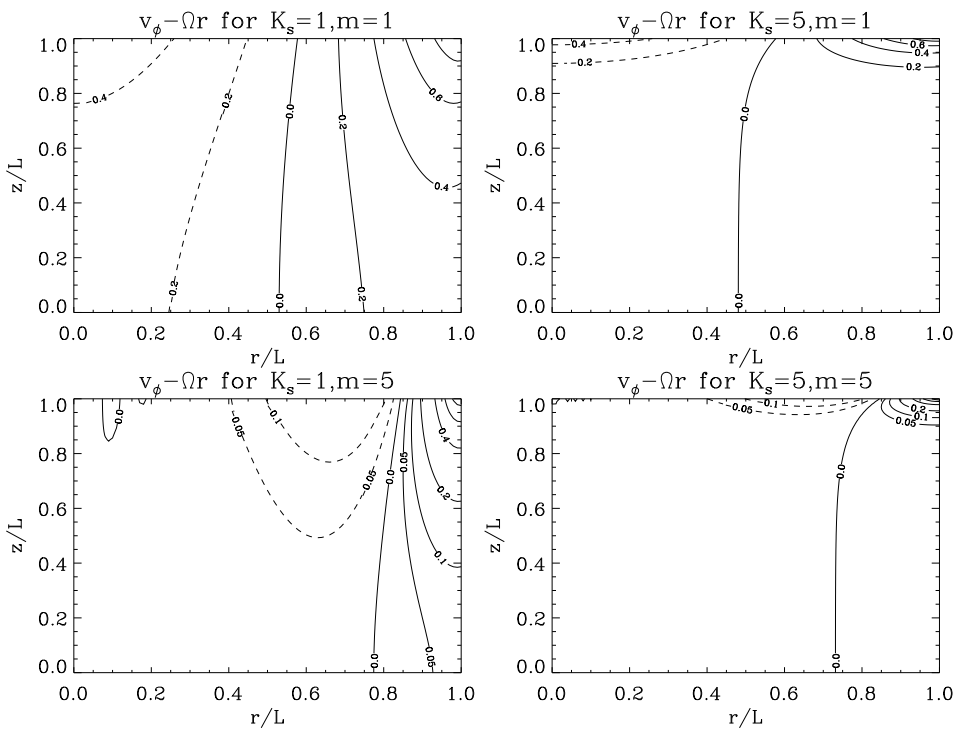

\begin{center}
  \includegraphics[scale=.4]{strat1a.epsi}
  \includegraphics[scale=.4]{strat1b.epsi} \\
  \includegraphics[scale=.4]{strat1c.epsi}
  \includegraphics[scale=.4]{strat1d.epsi}
\caption{Steady-state azimuthal velocity contours in the meridional plane $\phi-\Omega t=0$ in the rotating frame, illustrating the effects of stratification and mode number.  Contour values, top panels: $v_\phi-\Omega r=0.8, 0.6, 0.4, 0.2, 0$ (solid lines),$ -0.2, -0.4$ (dashed lines).  Bottom panels: $v_\phi-\Omega r=0.8, 0.6, 0.4, 0.2, 0.1, 0.05, 0$ (solid lines),$ -0.05, -0.1$ (dashed lines).  Stratification and mode number: $(K_s,m)=(1,1)$ (top left), $(5,1)$ (top right), $(1,5)$ (bottom left), $(5,5)$ (bottom right).  We set $F=1$ and $K=0.5$ in all panels.}
\end{center}
\label{fig1}
\end{figure}

\begin{figure}[h]
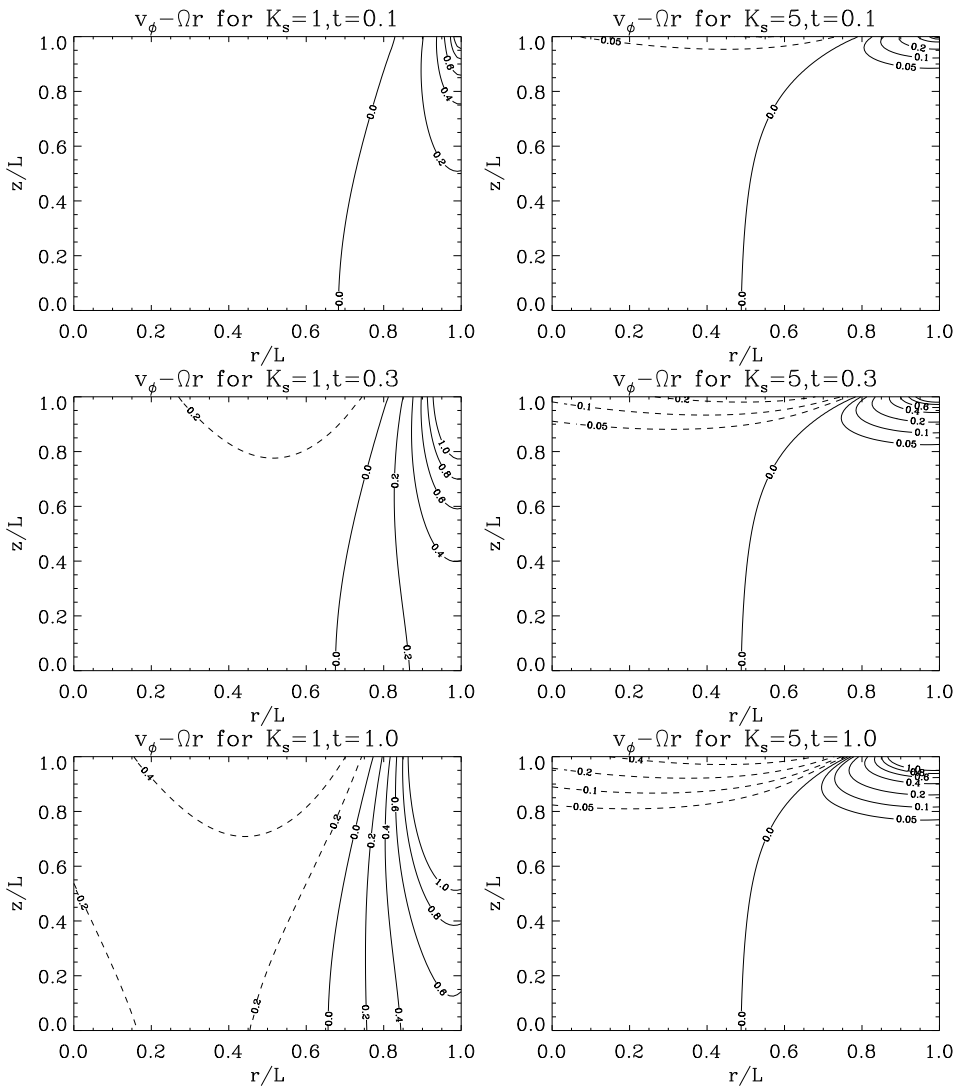

\begin{center}
   \includegraphics[scale=.4]{strat2a.epsi}
   \includegraphics[scale=.4]{strat2d.epsi} \\
   \includegraphics[scale=.4]{strat2b.epsi}
   \includegraphics[scale=.4]{strat2e.epsi} \\
   \includegraphics[scale=.4]{strat2c.epsi}
   \includegraphics[scale=.4]{strat2f.epsi} \\
\caption{Temporal evolution of azimuthal velocity contours in the meridional plane $\phi-\Omega t=0$ in the rotating frame, showing the effect of stratification.  Five modes $(1\leq m\leq 5)$, equally weighted, are included in the solution.  Contour values, left panels: $v_\phi-\Omega r=0.8, 0.6, 0.4, 0.2, 0$ (solid lines), $-0.2, -0.4 $(dashed lines).  Right panels: $v_\phi-\Omega r=1.0, 0.8, 0.6, 0.4, 0.2, 0.1, 0.05, 0$ (solid lines),$ -0.05, -0.1, -0.2, -0.4 $(dashed lines).  Each column contains three snapshots, at $tE^{1/2}\omega_{11}=0.1$ (top), $0.3$ (middle), and $1.0$ (bottom). Stratification parameter: $K_s=1$ (left),$ 5$ (right).  We set $F=1$ and $K=0.5$ in all panels.}
\end{center}
\label{fig2}
\end{figure}

Figure 1 summarizes the above physics pictorially.  The four panels display contour plots of $v_\phi-\Omega r$ at $t=E^{-1/2} \omega_{m1}^{-1}$ in the meridional plane $\phi-\Omega t=0$, for modes $m=1,5$ (top, bottom), stratification $K_s=1,5$ (left, right), and fixed compressibility $K=0.5$.  In the right-hand panels, where $\partial \rho/\partial z$ is larger, the flow on the midplane lags the flow at $z=\pm 1$ by more than in the left-hand panels, where $\partial \rho/\partial z$ is smaller.  The effect is more pronounced in the bottom panels, where $m$ is larger than in the top panels.  

Figure 2 illustrates how Ekman pumping proceeds with time when five modes $(1\leq m \leq 5)$ are included with equal weighting in (\ref{eq:gligw39}).  The left-hand panels are for $K_s=1$ and correspond to snapshots at $t E^{1/2} \omega_{11}=0.1,0.3,1.0$ running down the page.  The right-hand panels show the same thing for $K_s=5$.  Again, as time passes, the Ekman layer spreads from the top and sides into the interior of the cylinder, but it does so less rapidly when the hydrostatic density gradient is greater.  The time for $v_\phi-\Omega r$ to reach $e^{-1}$ of its maximum value at $z=\pm 1$ and $r=0$ in the left and right-hand panels is $t=0.53$ and $0.50$ respectively (in dimensionless units).

\section{Compressibility
 \label{sec:gligw4}}
Compressibility acts in two ways. (1) It reduces the buoyancy force experienced by an element of fluid displaced from its equilibrium position in the hydrostatic density gradient.  It therefore competes against the density gradient to increase both the spin-up time and the volume of fluid spun up.  (2) It also affects the spin-up time through mass conservation.  As the axial secondary flow moves from the top (or bottom) face to the midplane through the density gradient, it is compressed, decelerating the radial secondary flow and increasing the spin-up time. Therefore compressibility increases the spin-up time even when there is no buoyancy force.  This is demonstrated in the zero buoyancy limit $N\rightarrow0$, where we have $K_s=K$, $\beta_+=K_s$, $\beta_-=0$, and hence

\begin{eqnarray}
 v_r(r,\phi,z,t)= & 
 \frac{L \delta\Omega}{2} (1-e^{-E^{1/2}\omega t} ) \nonumber \\ 
 & \times \sum_{m=0}^\infty \left( \frac{r}{L}\right)^{m-1} \left[\left( \frac{r}{L}\right)^2-1 \right]  m \sin{m(\phi-\Omega t)}~,
 \label{eq:gligw47}
\end{eqnarray}
\begin{eqnarray}
 v_\phi(r,\phi,z,t)= &r \Omega +
 \frac{L \delta\Omega}{2} (1-e^{-E^{1/2}\omega t} ) \nonumber \\
 & \times \sum_{m=0}^\infty  \left( \frac{r}{L}\right)^{m-1} \left[(m+2)\left( \frac{r}{L}\right)^2 -m \right] \cos{m(\phi-\Omega t)}~,
 \label{eq:gligw48}
\end{eqnarray}
\begin{eqnarray}
 v_z(r,\phi,z,t)
 & = &
 E^{1/2}L \delta\Omega \left(\frac{e^{g z / c^2}-1}{e^{g L / c^2}-1}\right) e^{-E^{1/2}\omega t} 
 \nonumber \\
 & &
 \times \sum_{m=0}^\infty \left( \frac{r}{L}\right)^{m}\left( m+1\right)\cos{m(\phi-\Omega t)} ~,
 \label{eq:gligw49}
\end{eqnarray}
\begin{eqnarray}
 \rho(r,\phi,z,t)
 & = &
 \rho_{\rm 0} e^{-g z / c^2} + \frac{\rho_{\rm 0} L^2 \Omega \delta\Omega}{c^2} e^{-g z / c^2} ( 1-e^{-E^{1/2}\omega t}) 
 \nonumber \\
 & &
 \times 
\sum_{m=0}^\infty \left( \frac{r}{L}\right)^{m}\left[ \left( \frac{r}{L}\right)^2-1\right] \cos{m(\phi-\Omega t)}~,
 \label{eq:gligw50}
\end{eqnarray}
\begin{eqnarray}
 p(r,\phi,z,t)
 & = &
  \rho_{\rm 0} c^2 e^{-g z / c^2} + \rho_{\rm 0} L^2 \Omega \delta\Omega e^{-g z / c^2} (1-e^{-E^{1/2}\omega t} ) 
\nonumber \\
 & &
 \times \sum_{m=0}^\infty \left( \frac{r}{L}\right)^{m}\left[ \left( \frac{r}{L}\right)^2-1\right] \cos{m(\phi-\Omega t)}~.
 \label{eq:gligw51}
\end{eqnarray}
Note that $v_r$ and $v_\phi$ are independent of $z$ in this limit.  The reciprocal of the spin-up time is \cite{abn96} 
\begin{equation}
 \frac{\omega}{\Omega}
 = \frac{K}{e^{K}-1}~.
 \label{eq:gligw52}
\end{equation}
When there is no buoyancy, the spin-up time increases exponentially with the compressibility. All mode numbers $m$ are equally affected.
In the limit $c\rightarrow \infty$, the nonaxisymmetric generalisation of Ref. \cite{gre63} is obtained.

\begin{figure}[h]
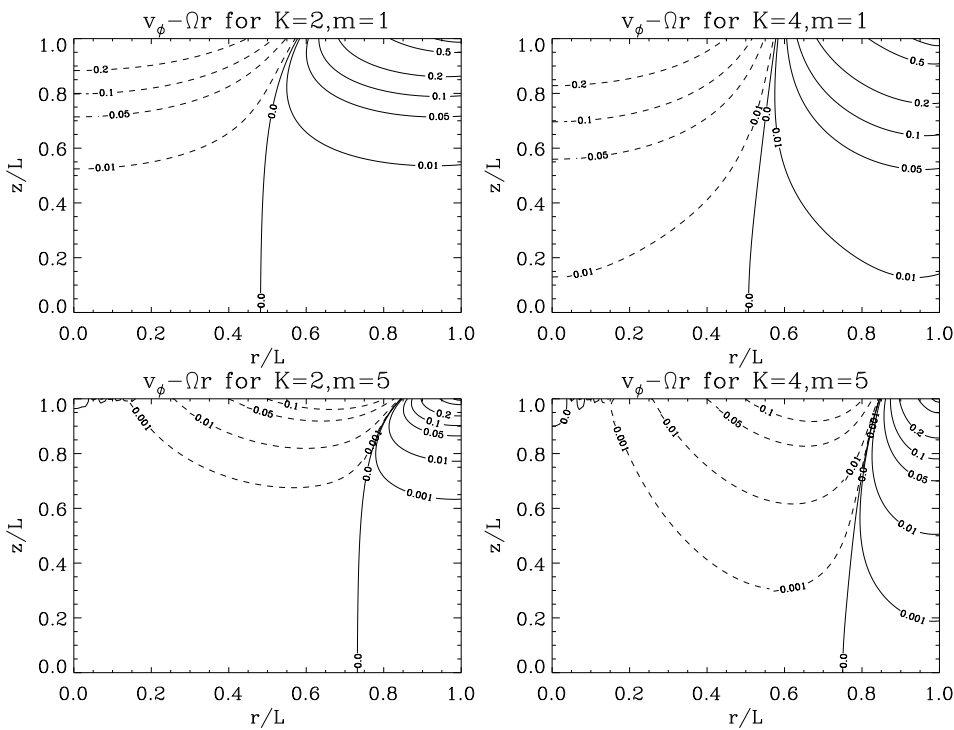

\begin{center}
  \includegraphics[scale=.4]{comp1a.epsi}
  \includegraphics[scale=.4]{comp1b.epsi} \\
  \includegraphics[scale=.4]{comp1c.epsi}
  \includegraphics[scale=.4]{comp1d.epsi}
\caption{Steady-state azimuthal velocity contours in the meridional plane $\phi-\Omega t=0$ in the rotating frame, illustrating the effects of compressibility and mode number.  Contour values, top panels: $v_\phi-\Omega r=0.5, 0.2, 0.1, 0.05, 0.01, 0$ (solid lines), $-0.01, -0.05, -0.1, -0.2$ (dashed lines).  Bottom panels: $v_\phi-\Omega r=0.5, 0.2, 0.1, 0.05, 0.01, 0.001, 0$ (solid lines), $-0.001, -0.01, -0.05, -0.1, -0.2$ (dashed lines).  Compressibility and mode number: $(K,m)=(2,1)$ (top left), $(4,1)$ (top right), $(2,5)$ (bottom left), $(4,5)$ (bottom right).  We set $F=1$ and $K=0.5$ in all panels.}
\end{center}
\label{fig3}
\end{figure}

\begin{figure}[h]
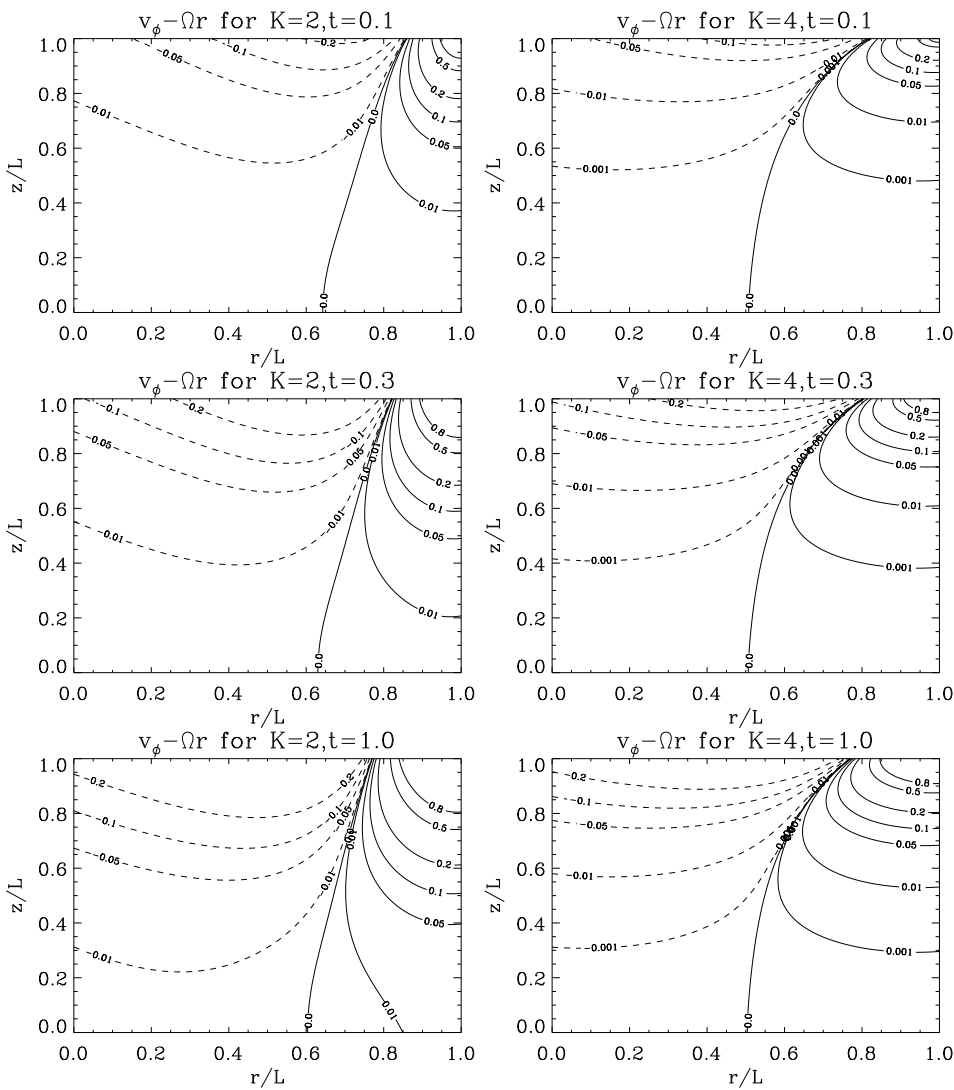

\begin{center}
   \includegraphics[scale=.4]{comp2a.epsi}
   \includegraphics[scale=.4]{comp2d.epsi} \\
   \includegraphics[scale=.4]{comp2b.epsi}
   \includegraphics[scale=.4]{comp2e.epsi} \\
   \includegraphics[scale=.4]{comp2c.epsi}
   \includegraphics[scale=.4]{comp2f.epsi} \\
\caption{Temporal evolution of azimuthal velocity contours in the meridional plane $\phi-\Omega t=0$ in the rotating frame, showing the effect of compressibility.  Five modes $(1\leq m\leq 5)$, equally weighted, are included in the solution.  Contour values, left panels: $v_\phi-\Omega r=0.8, 0.5, 0.2, 0.1, 0.05, 0.01, 0$ (solid lines), $-0.01, -0.05, -0.1, -0.2$ (dashed lines).  Right panels: $v_\phi-\Omega r=0.8, 0.5, 0.2, 0.1, 0.05, 0.01, 0.001, 0$ (solid lines), $-0.001, -0.01, -0.05, -0.1, -0.2$ (dashed lines).  Each column contains three snapshots, at $tE^{1/2}\omega_{11}=0.1$ (top), $0.3$ (middle), and $1.0$ (bottom). Compressibility parameter: $K=2$ (left), $4$ (right) We set $F=1$ and $K_s=5$ in all panels.}
\end{center}
\label{fig4}
\end{figure}

Figure 3 summarizes the above physics pictorially.  The four panels display contour plots of $v_\phi-\Omega r$ at $t=E^{-1/2} \omega_{m1}^{-1}$ in the meridional plane $\phi-\Omega t=0$, for modes $m=1,5$ (top, bottom), compressibility $K=2,4$ (left, right), and fixed stratification $K_s=5$.  In the left-hand panels, where $c^2$ is larger, the flow on the midplane lags the flow at $z=\pm 1$ by more than in the right-hand panels, where $c^2$ is smaller.  The effect is more pronounced on the bottom panels, where $m$ is larger than in the top panels.  

Figure 4 illustrates how Ekman pumping proceeds with time when five modes $(1\leq m \leq 5)$ are included with equal weighting in (\ref{eq:gligw39}).  The left-hand panels are for $K=2$ and correspond to snapshots at $t E^{1/2} \omega_{11}=0.1,0.3,1.0$ running down the page.  The right-hand panels show the same thing for $K=4$.  Again, as time passes, the Ekman layer spreads from the top and sides into the interior of the cylinder, but it does so more slowly when the compressibility is greater.  The time for $v_\phi-\Omega r$ to reach $e^{-1}$ of its maximum value at $z=\pm 1$ and $r=0$ in the left and right-hand panels is $t=0.55$ and $0.63$ respectively (in dimensionless units).

\section{Gravitational wave signal
 \label{sec:gligw5}}

The nonaxisymmetric secondary (spin-up) flow creates a stress energy distribution with a time-varying mass quadrupole moment, which emits gravitational radiation.  In this section, we calculate the gravitational wave signal from the flow solution derived in \S\ref{sec:gligw2}, for the special cases of polar and equatorial observers.  These special cases suffice to demonstrate, in principle, how the signal can be inverted to infer the internal properties of the star, specifically its viscosity and compressibility, as described in \S\ref{sec:gligw6}. A more general treatment, valid for arbitrary inclination angles and including the current quadrupole moment, is postponed to a future paper.

\subsection{Stress-energy tensor}

The relativistic stress-energy tensor of a perfect fluid is given by
\begin{equation}
  T^{\mu\nu}=\left( \rho+\frac{p}{c^2}\right)u^\mu u^\nu-p g^{\mu\nu}~,
 \label{eq:gligw53}
\end{equation}
where $u^\mu$ is the fluid 4-velocity and $g^{\mu \nu}$ is the metric tensor.  We neglect viscous terms in (\ref{eq:gligw53}) as they are of order $O(E)$.  In the quadrupole moment formalism, the gravitational wave signal is proportional to $T^{00}$.  Expanding $T^{00}$ in powers of $v/c$, expanding $\vec{v}$, $\rho$ and $p$ in powers of $\epsilon$ as in \S\ref{sec:gligw2}, and noting that we are now working in the inertial frame of the observer, we obtain
\begin{equation}
  T^{00}= \rho c^2+\epsilon\left( c^2\delta\rho + 2\rho \vec{v} \cdot \delta \vec{v}+2 \frac{p}{c^2} \vec{v} \cdot \delta \vec{v}\right)~.
 \label{eq:gligw54}
\end{equation}
upto and including terms of order $O(\epsilon)$ and $O(v/c)$.  Terms in (\ref{eq:gligw54}) containing only equilibrium quantities are static and axisymmetric and do not contribute to the gravitational wave signal.  The remaining terms are in the ratios 
\begin{eqnarray}
  \delta \rho c^2  : 2 \rho \left| \vec{v} \cdot \delta \vec{v} \right|:2 \frac{p}{c^2}\left| \vec{v} \cdot \delta \vec{v} \right| = 1:\left(\frac{g L }{c^2}\right):\left(\frac{g L }{c^2}\right)^2
 \label{eq:gligw55}
\end{eqnarray}
Therefore in the limit where the centrifugal force is much smaller than the Coriolis force, the leading order terms are

\begin{equation}
  T^{00}=\delta\rho c^2 +2  \rho \vec{v} \cdot \delta \vec{v}+2 \frac{p}{c^2} \vec{v} \cdot \delta \vec{v}
 \label{eq:gligw56}
\end{equation}
and $T^{00}$ is of characteristic magnitude $\Omega \delta \Omega L \rho_0 c^2 /g$.
If $gL/c^2$ is also small, as in a neutron star, then the dominant stress-energy contribution is $\delta \rho c^2$, with $\delta\rho $ given by (\ref{eq:gligw44}).  Note that, from equation (\ref{eq:gligw53}) onwards, $c$ now denotes the speed of light, not the sound speed (cf. \S\ref{sec:gligw2a}).

\subsection{Wave strain for a polar observer \label{sec:gligw5a}}
In Einstein's quadrupole moment formalism, the components of the gravitational wave strain in the transverse traceless gauge for an observer positioned along the $z'$ axis are 
\begin{eqnarray} \label{zstrains}
h_{x'x'}^{\rm TT}(t) & = & -h_{y'y'}^{\rm TT}(t)=\frac{G}{c^4d} \left[\ddot{I}_{x'x'}(t)-\ddot{I}_{y'y'}(t)\right] \label{eq:gligw57} \\
h_{x'y'}^{\rm TT}(t) & = &  \frac{2 G}{c^4 d}\ddot{I}_{x'y'}(t) \label{eq:gligw58}
\end{eqnarray}
where an overdot denotes a time derivative, $d$ is the distance to the source, and $I_{ij}$ is the reduced quadrupole moment-of-inertia tensor defined by
\begin{equation}
I_{ij}(t)=\frac{1}{c^2}\int d^3 \vec{x} \left( x_i x_j -\delta_{ij}\frac{\left|\vec{x}\right|^2}{3}\right)T^{00}(\vec{x},t)
 \label{eq:gligw59}~.
\end{equation}

Consider a polar observer located along the rotation ($z$) axis.  Equations (\ref{eq:gligw57})--(\ref{eq:gligw59}) can be combined with (\ref{eq:gligw44}) to give the following wave strains in the plus and cross polarisations respectively:

\begin{eqnarray}
h_{(\rm P)+}(t) = & \frac{h_0}{\Omega^2} \sum_{n=1}^\infty \alpha_{2n} \mathcal{Z}_{2n} \left\{ \left[\left(4\Omega^2-E \omega_{2n}^2\right)e^{-E^{1/2}w_{2n}t}-4 \Omega^2 \right]\cos{2\Omega t} \right. \nonumber \\
& \left. - 4 E^{1/2} w_{2n}\Omega e^{-E^{1/2}w_{2n}t}\sin{2 \Omega t} \right\}~,
 \label{eq:gligw60}
\end{eqnarray}
\begin{eqnarray}
h_{(\rm P)\times}(t) = & \frac{h_0}{\Omega^2}  \sum_{n=1}^\infty  \alpha_{2n} \mathcal{Z}_{2n} \left\{  4 E^{1/2} w_{2n}\Omega e^{-E^{1/2}w_{2n}t} \cos{2\Omega t} \right. \nonumber \\
& \left. + \left[(4\Omega^2-E \omega_{2n}^2)e^{-E^{1/2}w_{2n}t}-4\Omega^2\right] \sin{2 \Omega t} \right\} 
 \label{eq:gligw61}~.
\end{eqnarray}
In (\ref{eq:gligw60}) and (\ref{eq:gligw61}), we define the dimensionless characteristic amplitude 
\begin{eqnarray}
h_0= \left(\frac{\delta \Omega}{\Omega}\right)\left(\frac{\pi \rho_0 \Omega^4 G L^6}{c^4 g d}\right)~,
 \label{eq:gligw61b}
\end{eqnarray}
and the coefficients $\alpha_{mn}=A_{mn} J_{m+1}(\lambda_{mn})/\lambda_{mn}$, the first few of which are $\alpha_{11}=-0.074$, $\alpha_{12}=-0.007$, $\alpha_{21}=-0.035$, $\alpha_{22}=0.005$.  The factors $\mathcal{Z}_{mn}$ are also dimensionless, with
\begin{eqnarray}
\mathcal{Z}_{mn}= & 2 \left[(F N^2-\beta_-) e^{\beta_+}-(F N^2-\beta_+)e^{\beta_-}\right]^{-1} \nonumber \\
 & \times \int_{0}^{1} dz\left[ (F N^2-\beta_-)\beta_- e^{-\beta_- z}-(F N^2 - \beta_+)\beta_+e^{-\beta_+ z} \right] z^{2-m}~.
 \label{eq:gligw62}
\end{eqnarray}

In \S\ref{sec:gligw6}, we present a simple recipe for relating features of the gravitational wave spectrum to the physical properties of the flow.  We do this for the Fourier spectrum (instead of the raw signal in the time domain) to facilitate presentation of the physics; in practice, a time-domain template would be used to analyse real gravitational wave data.  Noting that $\alpha_{2n}$ diminish quickly with increasing $n$, we can approximate (\ref{eq:gligw60}) and (\ref{eq:gligw61}) by the first terms in their series.  The moduli of the Fourier transformed cross and plus wave strains are then
\begin{eqnarray}
\frac{|\bar{h}_{(\rm P)+}(\omega)|^2 }{h_0^2 \alpha^2_{21} |\mathcal{Z}_{21}|^2/\Omega^4}& = &   \frac{8\Omega^2E \omega_{21}^2(E \omega_{21}^2-\omega^2)+(16 \omega^4+E^2 \omega_{21}^4)(E \omega_{21}^2+\omega^2)}{(4\Omega^2+E \omega_{21}^2-\omega^2)^2+(2\omega E^{1/2}\omega_{21} )^2} ~,
 \label{eq:gligw63}
\end{eqnarray}
\begin{eqnarray}
\frac{|\bar{h}_{(\rm P)\times}(\omega)|^2  }{h_0^2 \alpha^2_{21} |\mathcal{Z}_{21}|^2/\Omega^4} =   \frac{2 \Omega\left[ 16 \omega^4 +4(2\Omega^2+\omega^2)E \omega_{21}^2+E^2 \omega_{21}^4\right]}{(4\Omega^2+E \omega_{21}^2-\omega^2)^2+(2\omega E^{1/2}\omega_{21} )^2}~.
 \label{eq:gligw64}
\end{eqnarray}
Equations (\ref{eq:gligw63}) and (\ref{eq:gligw64}) describe Lorentzian spectra, as for a damped harmonic oscillator with resonances at $\omega^2=4\Omega^2+E \omega_{21}^2$ and damping coefficient $E^{1/2}\omega_{21}$.  In neutron stars, which have $E^{1/2}\omega_{21}\ll 2 \Omega$, the Fourier peaks are much narrower than their separation, and hence we can safely ignore the interference (product) terms between them:
\begin{eqnarray}
|\bar{h}_{(\rm P)+}(\omega\simeq2\Omega)| = |\bar{h}_{(\rm P)\times}(\omega\simeq2\Omega)| = \frac{2 h_0 \alpha_{21} |\mathcal{Z}_{21}| }{\sqrt{(2 \Omega-\omega)^2 +E\omega_{21}^2} } 
 \label{eq:gligw65}
\end{eqnarray}
The spectra of the two polarizations are therefore indistinguishable for a polar observer.  

\begin{figure}[h]
\begin{center}
  \includegraphics[scale=.4]{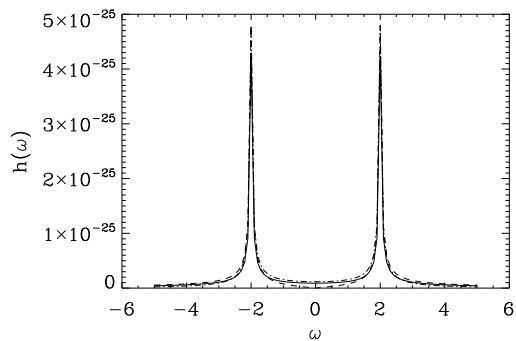}
\caption{Full plus and cross polarization spectra for a polar observer, compared with the approximate result for $E=10^{-4}$ and $K$, $N = 1$.  Dashed curve: Fourier transform of (\ref{eq:gligw60}); dashed-dotted curve: Fourier transform of (\ref{eq:gligw60}); solid curve: approximate result (\ref{eq:gligw65}).  The sum is taken to 50 terms in (\ref{eq:gligw60}) and (\ref{eq:gligw61}).}
\end{center}
\label{fig5}
\end{figure}

For small values of $E$, the approximate result (\ref{eq:gligw65}) agrees with the exact Fourier transform of (\ref{eq:gligw60}) and (\ref{eq:gligw61}) (with terms $1 \leq n \leq 50$ included) to an accuracy of $4\%$.  A comparison is displayed in Figure 5, where (\ref{eq:gligw65}) is plotted together with the Fourier transform of (\ref{eq:gligw60}) and (\ref{eq:gligw61}) for $E=10^{-4}$ and $K$, $N = 1$.

\subsection{Wave strain for an equatorial observer}
For an equatorial observer located along the $x$-axis, the plus and cross wave strains as functions of time are given by
\begin{eqnarray}
h_{(\rm E)+}(t) = & \frac{h_0}{2 \Omega^2} \sum_{n=1}^\infty \alpha_{2n} \mathcal{Z}_{2n} \left\{ \left[\left(4\Omega^2-E \omega_{2n}^2\right)e^{-E^{1/2}w_{2n}t}-4 \Omega^2 \right]\cos{2\Omega t}\right. \nonumber \\
& \left. - 4 E^{1/2} w_{2n}\Omega e^{-E^{1/2}w_{2n}t} \sin{2 \Omega t} \right\}~,
 \label{eq:gligw66}
\end{eqnarray}
\begin{eqnarray}
h_{(\rm E)\times}(t) = & \frac{2 h_0}{\Omega^2} \sum_{n=1}^\infty  \alpha_{1n} \mathcal{Z}_{1n}   \left\{ [2 E^{1/2} w_{1n}\Omega e^{-E^{1/2}w_{1n}t}\cos{\Omega t} \right. \nonumber \\
& \left. + \left[\left(\Omega^2-E \omega_{1n}^2\right)e^{-E^{1/2}w_{2n}t}-\Omega^2\right] \sin{\Omega t} \right\}~.
 \label{eq:gligw67}
\end{eqnarray}
From (\ref{eq:gligw60}) and (\ref{eq:gligw66}), it can be seen that the plus polarisation for an equatorial observer is exactly half as strong as the plus polarisation for a polar observer (and $\pi$ radians out of phase).  In addition, the cross polarisation for an equatorial observer oscillates at $\Omega$ rather than $2 \Omega$.  Proceeding as in \S\ref{sec:gligw5a}, we obtain the following ($n=1$) approximations for the plus and cross spectra:
\begin{eqnarray}
|\bar{h}_{(\rm E)+}(\omega)| = \frac{ h_0 \alpha_{21} |\mathcal{Z}_{21}|}{\sqrt{(2 \Omega-\omega)^2 +E\omega_{21}^2} }~,
 \label{eq:gligw68}
\end{eqnarray}
\begin{eqnarray}
|\bar{h}_{(\rm E)\times}(\omega)| = \frac{ h_0 \alpha_{11} |\mathcal{Z}_{11}|}{\sqrt{(\Omega-\omega)^2 +E\omega_{11}^2} } ~.
 \label{eq:gligw69}
\end{eqnarray}

\section{Measuring the viscosity and compressibility of bulk nuclear matter
 \label{sec:gligw6}}

The detailed wave form computed in \S\ref{sec:gligw5} can be inverted to measure $N$, $E$, and $K_s$ (and hence $K$, given $F$.  

To illustrate how, let us consider first a polar observer.  For such an observer, the gravitational wave spectrum peaks at $\omega=2 \Omega$.  The amplitude of the peak is
\begin{equation}
M\left[h_{(\rm P)+}\right]=M\left[h_{(\rm P)\times}\right]= 2  h_0 \alpha_{21} |\mathcal{Z}_{21}| E^{-1/2} \omega_{21}^{-1}
\label{eq:gligw70}
\end{equation}
and its width is
\begin{equation}
\Gamma\left[h_{(\rm P)+}\right]=\Gamma\left[h_{(\rm P)\times}\right]=\sqrt{3}E^{1/2} \omega_{21}~.
\label{eq:gligw71}
\end{equation}
A more useful expression than (\ref{eq:gligw70}) is the product of the peak amplitude and width
\begin{equation}
M\left[h_{(\rm P)+}\right]\Gamma\left[h_{(\rm P)+}\right]=M\left[h_{(\rm P)\times}\right]\Gamma\left[h_{(\rm P)\times}\right]=2 \sqrt{3} h_0 \alpha_{21} |\mathcal{Z}_{21}| ,
\label{eq:gligw72}
\end{equation}
which is independent of $E$.  If the distance to the source is somehow known, along with $\delta \Omega$ (e.g. from radio timing data), then $h_0$ is known.  If, in addition we can develop a reliable estimate of $\nu$ and hence $E$, 
(\ref{eq:gligw71}) and (\ref{eq:gligw72}) can be solved for $\omega_{21}$ and $|\mathcal{Z}_{21}|$.  As these quantities are functions of $K_s$ and $N$, (\ref{eq:gligw71}) and (\ref{eq:gligw72}) constrain the internal properties of the neutron star.  If $h_0$ and $E$ are unknown, then (\ref{eq:gligw71}) and (\ref{eq:gligw72}) place constraints on the latter two quantites, as well as $K_s$ and $N$.

\begin{figure}[h]
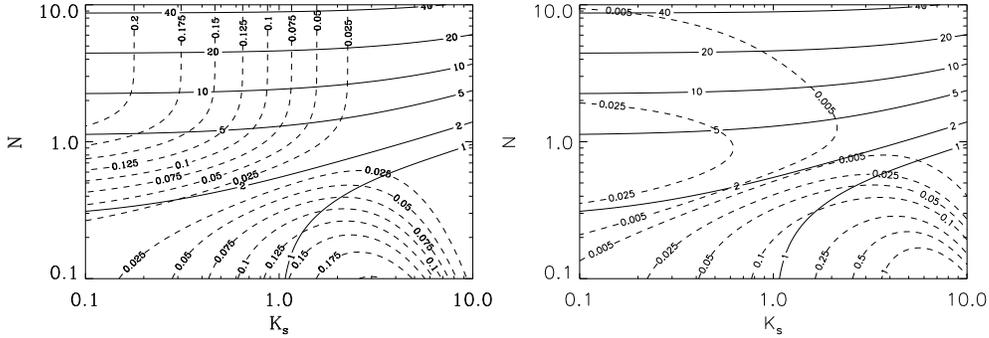

\begin{center}
  \includegraphics[scale=.4]{axplot.epsi}
  \includegraphics[scale=.4]{axplot2.epsi}
\caption{Width and amplitudes for the gravitational wave spectrum seen by a polar observer, showing the effect of buoyancy and stratification.  Two sets of contours are plotted in each panel. Left: width (\ref{eq:gligw71}), (solid curve),  width $\times$ amplitude (\ref{eq:gligw72}), (dashed curve).  Contour values: $\Gamma E^{-1/2}=1, 2, 5, 10, 20, 40$ and $\Gamma M/ h_0=0.025, 0.05, 0.075, 0.125, 0.15, 0.175, 0.2$. Right: width (\ref{eq:gligw71}), (solid curve), amplitude (\ref{eq:gligw70}), (dashed).  Contour values: $\Gamma E^{-1/2}=1,2,5,10,20,40$ and $\Gamma/ h_0= 0.005, 0.025, 0.05, 0.1, 0.25, 0.5, 1$.}
\end{center}
\label{fig6}
\end{figure}

Figure 6 presents two panels.  On the left, contours of constant $\Gamma\left[h_{(\rm P)+}\right] E^{-1/2}$ and constant $\Gamma\left[h_{(\rm P)+}\right] M\left[h_{(\rm P)+}\right]/ h_0$ are plotted as solid and dashed curves as a function of buoyancy and stratification.  On the right, contours of constant $\Gamma\left[h_{(\rm P)+}\right] E^{-1/2}$ and constant $M\left[h_{(\rm P)+}\right]/ h_0$ are plotted.  The figure illustrates how measurement of the width and amplitude of the gravitational wave spectrum for a polar observer can be used to constrain the quantities $\Gamma E^{-1/2}$ and $\Gamma M/ h_0$.  In addition, if $E$ and $h_0$ are known, the intersection of the measured solid and dashed contours represents a direct measurement of the properties of bulk nuclear matter.  The width of the spectral peak depends most on the buoyancy (nearly horizontal contours).  In contrast, the amplitude of the spectral peak exhibits a saddle at $(K_s,N)\sim(1,0.4)$, corresponding to an inversion in sign of the strain.  For high buoyancies, $\Gamma\left[h_{(\rm P)+}\right] M\left[h_{(\rm P)+}\right]$ depends only on stratification.

Although Figure 6 can be used in principle to infer the internal properties of the neutron star from the width and amplitude of the gravitational wave spectrum, it is rare in practice that $h_0$ and $E$ are known.  For the special case of a polar observer, the plus and cross polarisations are identical, and all the gravitational radiation is emitted at $\omega=2\Omega$, so we only have two measurable parameters, the width and amplitude of the spectral peak at $\omega=2\Omega$, to constrain the unknowns $K,N,h_0$ and $E$.
As we show below, an equatorial observer receives radiation at $\omega=\Omega$ and $\omega=2 \Omega$ and thereby has four measurable parameters at his disposal.  However, an observer at an inclination angle $0<i<\pi/2$ observes signals at $\Omega$ and $2\Omega$ with $h_{+}\neq h_{\times}$ at both frequencies.  Therefore, such an observer has eight independent peak amplitudes and widths to play with --- enough, indeed, to solve uniquely for $K,N,h_0$ and $E$, as well as the inclination angle, $i$.  This general case, which is more complicated, will be investigated in a forthcoming paper \cite{ben08}.

For an equatorial observer, the $h_{+}$ spectrum peaks at $2\Omega$, while the $h_{\times}$ spectrum peaks at $\Omega$.  The respective widths and amplitudes are
\begin{equation}
\Gamma\left[h_{(\rm E)+}\right]=\sqrt{3}E^{1/2} \omega_{21}~,
\label{eq:gligw73}
\end{equation}
\begin{equation}
\Gamma\left[h_{(\rm E)\times}\right]=\sqrt{3}E^{1/2} \omega_{11}~,
\label{eq:gligw74}
\end{equation}
\begin{equation}
M\left[h_{(\rm E)+}\right]\Gamma\left[h_{(\rm E)+}\right]=\sqrt{3} h_0 \alpha_{21} |\mathcal{Z}_{21}|~,
\label{eq:gligw75}
\end{equation}
\begin{equation}
M\left[h_{(\rm E)\times}\right]\Gamma\left[h_{(\rm E)\times}\right]=\sqrt{3} h_0 \alpha_{11} |\mathcal{Z}_{11}| ~.
\label{eq:gligw76}
\end{equation}
Equations (\ref{eq:gligw73})--(\ref{eq:gligw76}) contain four measureable quantities, which can be used to constrain the unknown quantities $K,N,h_0$ and $E$.  

\begin{figure}[h]
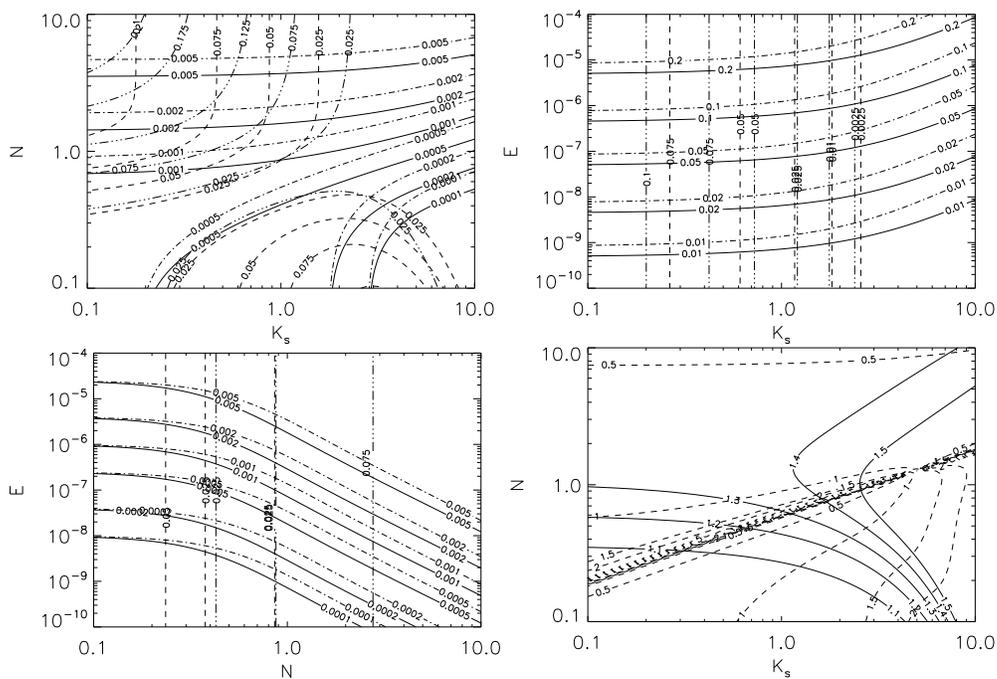

\begin{center}
  \includegraphics[scale=.4]{eqplotKN.epsi}
  \includegraphics[scale=.4]{eqplotKE.epsi} \\
  \includegraphics[scale=.4]{eqplotNE.epsi}
  \includegraphics[scale=.4]{eqplotcom.epsi} \\
\caption{Spectral line width and amplitude contours for the plus and cross polarisations seen by an equatorial observer, showing the variation with $N$, $K_s$ and $E$.  Top-left: $N$ versus $K_s$ for $E=10^{-7}$, $M\left[h_{(\rm E)+}\right] \Gamma\left[h_{(\rm E)+}\right]/h_0=0.025, 0.05, 0.075, 0.1$ (dashed contours),  $M\left[h_{(\rm E)\times}\right] \Gamma\left[h_{(\rm E)\times}\right]/h_0=0.025, 0.075, 0.125, 0.175, 0.2$ (dash-triple-dotted contours), $\Gamma\left[h_{(\rm E)+}\right]$ (solid contours)$=\Gamma\left[h_{(\rm E)\times}\right]$ (dash-dotted contours)$= 0.0001, 0.0002, 0.0005, 0.001, 0.002, 0.005$.  Top-right: $E$ versus $K_s$ for $N=1$,  $M\left[h_{(\rm E)+}\right] \Gamma\left[h_{(\rm E)+}\right]/h_0$ (dashed contours)=$M\left[h_{(\rm E)\times}\right] \Gamma\left[h_{(\rm E)\times}\right]/h_0$ (dash-triple-dotted contours)$=0.0025, 0.01, 0.025, 0.05, 0.075, 0.1$, $\Gamma\left[h_{(\rm E)+}\right]$ (solid contours) $=\Gamma\left[h_{(\rm E)\times}\right]$ (dash-dotted contours) $=0.0001, 0.0003, 0.001, 0.003, 0.01, 0.05$.  Bottom-left: $E$ versus $N$ for $K_s=1$, contours as for top-right.  Bottom-right; $M\left[h_{(\rm E)+}\right] \Gamma\left[h_{(\rm E)+}\right] / M\left[h_{(\rm E)\times}\right] \Gamma\left[h_{(\rm E)\times}\right]=0.5, 1, 1.5, 2, 5$ (dashed contours), $\Gamma\left[h_{(\rm E)+}\right] / \Gamma\left[h_{(\rm E)\times}\right]=1, 1.1, 1.2, 1.3, 1.4, 1.5$ (solid contours).

}
\end{center}
\label{fig7}
\end{figure}

Figure 7 plots the relationships in (\ref{eq:gligw73})--(\ref{eq:gligw76}) graphically. The top-left, top-right and bottom-left panels show contours of $M \Gamma/h_0$ and $\Gamma$ for both the cross and plus polarizations for different slices of parameter space.  Slices are taken at $E=10^{-7}$, $N=1$, and $K_s=1$ respectively.  These panels can be used to solve for the four unknown quantities $K_s$, $N$, $E$ and $h_0$ by locating the intersection of the contours corresponding to the measurements of $\Gamma$ and $M$.  
The bottom-right panel has contours of the ratios $M\left[h_{(\rm E)+}\right] \Gamma\left[h_{(\rm E)+}\right] / M\left[h_{(\rm E)\times}\right] \Gamma\left[h_{(\rm E)\times}\right]$ and $\Gamma\left[h_{(\rm E)+}\right] / \Gamma\left[h_{(\rm E)\times}\right]$.  Because the width and amplitude ratios are independent of $E$ and $h_0$ in our approximation, the intersection of contours in this panel allows a direct evaluation of $N$ and $K_s$.  The other panels can then be used to determine $E$ and $h_0$.  Therefore, observing a neutron star from the equator tells us more, in this context, than observing it from the pole.  Note that the results are sensitive to the measurement of the width ratio and amplitude ratio, since the widths and amplitudes of the two spectra depend on the stratification and buoyancy in a similar manner.

\section{Comparison with Ion Collider Experiments
 \label{sec:gligw8}}

The gravitational wave ``experiments'' proposed in \S\ref{sec:gligw6} to measure $K$, $N$, and $E$ dovetail with recent progress on measuring the compressibility and viscosity of bulk nuclear matter in terrestrial experiments. Indeed, it has been argued \cite{web07} that gravitational wave observations of pulsars will become the main experimental avenue for probing the nuclear equation of state in the future.

The compressibility is usually parameterised in terms of the compression modulus $\kappa$, which has the dimensions of energy; in our notation, we have $K=A m_p g L /\kappa$ (where $A$ is the mean atomic number, and $m_p$ is the proton mass).  It can be measured in one of two ways: (I) heavy-ion collision experiments, and (II) nuclear resonance experiments.  In (I), the ratio of $K^{+}$ meson multiplicities in $\rm Au+Au$ and $\rm C+C$ collisions (interpreted with the aid of nonequilibrium quantum molecular dynamics simulations) points to a soft equation of state, with $\kappa\approx200~\rm{MeV}$. \cite{stu01, fuc01, har06, ada07, for07} In (II), $\kappa$ is deduced from the centroid energies of the isoscalar giant-monopole and isovector giant-dipole resonances in heavy nuclei.  For $\rm{Zr}$ and $\rm{Pb}$, the data yield $243\,\rm{Mev}\leq\kappa\leq 268\,\rm{MeV}$, interpreted in the context of a relativistic mean-field theory of the nucleus. \cite{vre97, pie04, har06}  Again, a soft equation of state is preferred.  In our notation, (I) and (II) imply $0.73\leq K \leq0.97$.

The viscosity of bulk nuclear matter is usually parametrised in terms of the shear viscosity to entropy density ratio, $\eta/s$ (SI units: $\rm{K~s}$).  In our notation, we have $E=(A' k_B / m_p L^2 \Omega)(\eta/s)$ for the Ekman number, where $1 \leq A' \leq 2$ is the entropy per nucleon in units of Boltzmann's constant, $k_{\rm B}$.  The viscosity to entropy ratio $\eta/s$ can be measured in relativistic heavy-ion collision experiments, by looking at the percentage azimuthal anisotropy of the hadron flux released during the collision. \cite{adl03,ada07}  During the first $\sim 10^{-23}\,\rm{s}$ after the collision, the nucleons flow hydrodynamically (and elliptically), and $\eta/s$ is proportional to the resulting diffusion coefficient, which can be related to the azimuthal anisotropy.  Data from the latest PHENIX experiments ($\rm{Au+Au}$ collisions) imply $\eta/s\approx\hbar/4\pi k_{\rm B}$ for energies just above the quark deconfinement phase transition.  The latter value is the conjectured quantum lower bound for an ideal fluid.  It has been derived by exploiting the duality between a large class of gauge theories similar (but not identical) to quantum chromodynamics and the thermodynamics of black hole horizons. \cite{kov05,mat07}  In our notation, the measurements imply $E=5 \times 10^{-17} A' (L/10^4\,\rm m)^{-2} (\Omega/ \rm{rad\,s^{-1}})^{-1}$.  Interestingly, this is $\sim10$ orders of magnitude smaller than the Newtonian viscosity for neutron-neutron scattering in a superfluid calculated by Cutler and Lindblom. \cite{cut87,mel07b}

\section{Conclusions
 \label{sec:gligw9}}

In this paper, the gravitational wave signal emitted by the nonaxisymmetric spin-up flow excited by a pulsar glitch is calculated analytically in the context of a cylindrical, Newtonian-fluid toy model, which neglects the physics of pinning and turbulence in the boundary layer between the superfluid and the crust.  The model clarifies the physical roles played by stratification and compressibility in nonaxisymmetric spin up.  It is found that stratification and compressibility compete to increase and reduce the buoyancy respectively, and consequently, reduce and increase the spin-up time.  Higher order mode numbers enhance the effects of buoyancy.  Equations (\ref{eq:gligw65}), (\ref{eq:gligw68}) and (\ref{eq:gligw69}) give approximate formulas for the gravitational wave spectrum in the plus and cross polarizations.  Measurements of the amplitudes and widths of the lines in these spectra can be used to constrain $K_s$, $N$, and $E$, by reading off the intersection of the contours plotted in Figures 6 and 7.  

Current estimates of the internal properties of a typical neutron star imply that the characteristic wave strain (\ref{eq:gligw61b}) produced by a large ($\delta\Omega/\Omega\geq10^{-4}$) glitch evaluates roughly to $h_0\geq5\times10^{-26}$ for $L=10^4\,\rm{m}$, mass $M_*=M_{\odot}$, $g\approx GM_*/L^2$, $d=1\,\rm{kpc}$, and $f=\Omega/2\pi\geq0.2\,\rm{kHz}$.  While this estimate is crude, in view of the over-simplified model assumptions discussed in \S\ref{sec:gligw1} and \S\ref{sec:gligw2}, it suggests that a glitch with these parameters arguably approaches the threshold of detectability by a coherent, matched filter search ($1\%$ false alarm, $10\%$ false dismissal) with Advanced LIGO for an observation time of $T_{\rm{obs}}=14\,\rm{days}$.  Note that $T_{\rm{obs}}$ is limited by two factors: (1) the computational expense of a fully coherent search with matched filters, which is prohibitive for $T_{\rm{obs}}>14\,\rm{d}$ at present \cite{abb04}; and (2) the duration of the exponential relaxation phase following a glitch (i.e., the Ekman time-scale), which is observed to last from hours to a few weeks in radio pulsar timing experiments \cite{lyn00, mcc90, won01}.  Restriction (2) is lifted if the nonaxisymmetry persists even after the relaxation phase, e.g. due to inhomogeneous vortex pinning \cite{che88,war08}.  Of course, compressibility and stratification act to reduce $h_+$ and $h_\times$ below the fiducial estimate of $h_0$ given above, sometimes substantially; one has $1\leq N\leq10$ and $0.1\leq K_s\leq1$ for a typical neutron star \cite{rei92,abn96,cut87,mel07b}.  The above estimates apply to the largest glitches observed to date, with $\delta\Omega/\Omega\geq10^{-4}$ \cite{per07}.  A recent analysis of the observed glitch size and waiting-time distributions in individual pulsars suggests that future monitoring will reveal even larger glitches \cite{mel07}.  Once detected, the glitch signal will allow us to compare astronomical measurements of $K$ and $E$ with data from recent heavy-ion collider experiments.  The collider data prefer a soft nuclear equation of state and a viscosity close to the quantum lower bound for an ideal fluid, contradicting some neutron star models.  

Although our toy analytical model involves many unavoidable approximations (e.g. cylindrical geometry, uniform gravity, Newtonian fluid, idealized superfluid boundary layer), it nevertheless demonstrates \textit{in principle}  for how the gravitational wave signal from a pulsar glitch can be inverted to extract important information about neutron stars and bulk nuclear matter, independently from terrestrial experiments with particle accelerators.  Our preliminary detectability estimates offer grounds for hope that such gravitational wave observations are realistic in the medium term.  Of course, to apply these ideas in practice, the idealised model in this paper is inadequate and must be supplanted by large-scale numerical simulations, e.g. Refs \cite{per05, per06b}.

\section{Acknowledgements}
The authors wish to thank Greg Mendell and the anonymous referees for their helpful feedback on this manuscript.

\newpage

\bibliographystyle{unsrt}
\bibliography{glitchgw1}

\begin{thebibliography}{10}

\bibitem{man01}
R.~N. {Manchester}, A.~G. {Lyne}, F.~{Camilo}, J.~F. {Bell}, V.~M. {Kaspi},
  N.~{D'Amico}, N.~P.~F. {McKay}, F.~{Crawford}, I.~H. {Stairs}, A.~{Possenti},
  M.~{Kramer}, and D.~C. {Sheppard}.
\newblock {The Parkes multi-beam pulsar survey - I. Observing and data analysis
  systems, discovery and timing of 100 pulsars}.
\newblock 328:17--35, November 2001.

\bibitem{hob02}
G.~{Hobbs}.
\newblock {\em {Searches for and timing of radio pulsars}}.
\newblock PhD thesis, AA(Univ.~of Manchester, United Kingdowm.), 2002.

\bibitem{kra03}
A.~{Krawczyk}, A.~G. {Lyne}, J.~A. {Gil}, and B.~C. {Joshi}.
\newblock {Observations of 14 pulsar glitches}.
\newblock {\em Monthly Notices of the Royal Astronomical Society},
  340:1087--1094, April 2003.

\bibitem{jan06}
G.~H. {Janssen} and B.~W. {Stappers}.
\newblock {30 glitches in slow pulsars}.
\newblock {\em Astronomy \& Astrophysics}, 457:611--618, October 2006.

\bibitem{dal03}
S.~{Dall'Osso}, G.~L. {Israel}, L.~{Stella}, A.~{Possenti}, and E.~{Perozzi}.
\newblock {The Glitches of the Anomalous X-Ray Pulsar 1RXS J170849.0-400910}.
\newblock {\em Astrophysical Journal}, 599:485--497, December 2003.

\bibitem{kas03}
V.~M. {Kaspi} and F.~P. {Gavriil}.
\newblock {A Second Glitch from the ``Anomalous'' X-Ray Pulsar 1RXS
  J170849.0-4000910}.
\newblock {\em Astrophysical Journall}, 596:L71--L74, October 2003.

\bibitem{mck90}
J.~{McKenna} and A.~G. {Lyne}.
\newblock {PSR1737 - 30 and period discontinuities in young pulsars}.
\newblock {\em Nature}, 343:349--+, January 1990.

\bibitem{she96}
S.~L. {Shemar} and A.~G. {Lyne}.
\newblock {Observations of pulsar glitches}.
\newblock {\em Monthly Notices of the Royal Astronomical Society},
  282:677--690, September 1996.

\bibitem{ura99}
J.~O. {Urama} and P.~N. {Okeke}.
\newblock {Vela-size glitch rates in youthful pulsars}.
\newblock {\em Monthly Notices of the Royal Astronomical Society},
  310:313--316, December 1999.

\bibitem{wan00}
N.~{Wang}, R.~N. {Manchester}, R.~T. {Pace}, M.~{Bailes}, V.~M. {Kaspi}, B.~W.
  {Stappers}, and A.~G. {Lyne}.
\newblock {Glitches in southern pulsars}.
\newblock {\em Monthly Notices of the Royal Astronomical Society},
  317:843--860, October 2000.

\bibitem{per07}
C.~A. {Peralta}.
\newblock {\em {Superfluid spherical Couette flow and rotational irregularities
  in pulsars}}.
\newblock PhD thesis, AA(Univ.~of Melbourne, Australia.), 2007.

\bibitem{mel07}
A.~{Melatos}, C.~{Peralta}, and J.~S.~B. {Wyithe}.
\newblock {Avalanche Dynamics of Radio Pulsar Glitches}.
\newblock {\em Astrophysical Journal}, 672:1103--1118, January 2008.

\bibitem{lyn00}
A.~G. {Lyne}, S.~L. {Shemar}, and F.~G. {Smith}.
\newblock {Statistical studies of pulsar glitches}.
\newblock {\em Monthly Notices of the Royal Astronomical Society},
  315:534--542, July 2000.

\bibitem{and75}
P.~W. {Anderson} and N.~{Itoh}.
\newblock {Pulsar glitches and restlessness as a hard superfluidity
  phenomenon}.
\newblock {\em Nature}, 256:25--27, July 1975.

\bibitem{lin96}
B.~{Link} and R.~I. {Epstein}.
\newblock {Thermally Driven Neutron Star Glitches}.
\newblock {\em Astrophysical Journal}, 457:844--+, February 1996.

\bibitem{jon98}
P.~B. {Jones}.
\newblock {The origin of pulsar glitches}.
\newblock {\em Monthly Notices of the Royal Astronomical Society},
  296:217--224, May 1998.

\bibitem{rei92}
A.~{Reisenegger} and P.~{Goldreich}.
\newblock {A new class of g-modes in neutron stars}.
\newblock {\em Astrophysical Journal}, 395:240--249, August 1992.

\bibitem{per05}
C.~{Peralta}, A.~{Melatos}, M.~{Giacobello}, and A.~{Ooi}.
\newblock {Global Three-dimensional Flow of a Neutron Superfluid in a Spherical
  Shell in a Neutron Star}.
\newblock {\em Astrophysical Journal}, 635:1224--1232, December 2005.

\bibitem{and03}
N.~{Andersson}, G.~L. {Comer}, and R.~{Prix}.
\newblock {Are Pulsar Glitches Triggered by a Superfluid Two-Stream
  Instability?}
\newblock {\em Physical Review Letters}, 90(9):091101--+, March 2003.

\bibitem{mas05}
A.~{Mastrano} and A.~{Melatos}.
\newblock {Kelvin-Helmholtz instability and circulation transfer at an
  isotropic-anisotropic superfluid interface in a neutron star}.
\newblock {\em Monthly Notices of the Royal Astronomical Society},
  361:927--941, August 2005.

\bibitem{jon02}
D.~I. {Jones} and N.~{Andersson}.
\newblock {Gravitational waves from freely precessing neutron stars}.
\newblock {\em Monthly Notices of the Royal Astronomical Society},
  331:203--220, March 2002.

\bibitem{per06a}
C.~{Peralta}, A.~{Melatos}, M.~{Giacobello}, and A.~{Ooi}.
\newblock {Gravitational Radiation from Nonaxisymmetric Spherical Couette Flow
  in a Neutron Star}.
\newblock {\em Astrophysical Journall}, 644:L53--L56, June 2006.

\bibitem{per06b}
C.~{Peralta}, A.~{Melatos}, M.~{Giacobello}, and A.~{Ooi}.
\newblock {Transitions between Turbulent and Laminar Superfluid Vorticity
  States in the Outer Core of a Neutron Star}.
\newblock {\em Astrophysical Journal}, 651:1079--1091, November 2006.

\bibitem{and07}
N.~{Andersson}, T.~{Sidery}, and G.~L. {Comer}.
\newblock {Superfluid neutron star turbulence}.
\newblock {\em Monthly Notices of the Royal Astronomical Society},
  381:747--756, October 2007.

\bibitem{rez00}
V.~{Rezania} and M.~{Jahan-Miri}.
\newblock {The possible role of r-modes in post-glitch relaxation of the Crab
  pulsar}.
\newblock {\em Monthly Notices of the Royal Astronomical Society},
  315:263--268, June 2000.

\bibitem{and01}
N.~{Andersson} and G.~L. {Comer}.
\newblock {Probing Neutron-Star Superfluidity with Gravitational-Wave Data}.
\newblock {\em Physical Review Letters}, 87(24):241101--+, December 2001.

\bibitem{sed03}
D.~M. {Sedrakian}, M.~{Benacquista}, K.~M. {Shahabassian}, A.~A. {Sadoyan}, and
  M.~V. {Hairapetyan}.
\newblock {Gravitational Radiation from Fluctuations in Rotating Neutron
  Stars}.
\newblock {\em Astrophysics}, 46:445--454, October 2003.

\bibitem{sed06}
D.~M. {Sedrakian}, M.~V. {Hayrapetyan}, and M.~K. {Shahabasyan}.
\newblock {Gravitational radiation of slowly rotating neutron stars}.
\newblock {\em Astrophysics}, 49:194--200, April 2006.

\bibitem{don06}
P.~{Donati} and P.~M. {Pizzochero}.
\newblock {Realistic energies for vortex pinning in intermediate-density
  neutron star matter}.
\newblock {\em Physics Letters B}, 640:74--81, September 2006.

\bibitem{avo07}
P.~{Avogadro}, F.~{Barranco}, R.~A. {Broglia}, and E.~{Vigezzi}.
\newblock {Quantum calculation of vortices in the inner crust of neutron
  stars}.
\newblock {\em Physics Review C}, 75(1):012805--+, January 2007.

\bibitem{sha77}
J.~{Shaham}.
\newblock {Free precession of neutron stars - Role of possible vortex pinning}.
\newblock {\em Astrophysical Journal}, 214:251--260, May 1977.

\bibitem{che88}
K.~S. {Cheng}, D.~{Pines}, M.~A. {Alpar}, and J.~{Shaham}.
\newblock {Spontaneous superfluid unpinning and the inhomogeneous distribution
  of vortex lines in neutron stars}.
\newblock {\em Astrophysical Journal}, 330:835--846, July 1988.

\bibitem{sed99}
A.~{Sedrakian}, I.~{Wasserman}, and J.~M. {Cordes}.
\newblock {Precession of Isolated Neutron Stars. I. Effects of Imperfect
  Pinning}.
\newblock {\em Astrophysical Journal}, 524:341--360, October 1999.

\bibitem{eas79}
I.~{Easson}.
\newblock {Postglitch behavior of the plasma inside neutron stars}.
\newblock {\em Astrophysical Journal}, 228:257--267, February 1979.

\bibitem{abn96}
M.~{Abney} and R.~I. {Epstein}.
\newblock {Ekman pumping in compact astrophysical bodies}.
\newblock {\em Journal of Fluid Mechanics}, 312:327--340, 1996.

\bibitem{vre97}
D.~{Vretenar}, G.~A. {Lalazissis}, R.~{Behnsch}, W.~{P{\"o}schl}, and
  P.~{Ring}.
\newblock {Monopole giant resonances and nuclear compressibility in
  relativistic mean field theory}.
\newblock {\em Nuclear Physics A}, 621:853--878, February 1997.

\bibitem{gle00}
N.~{Glendenning}.
\newblock {\em {Compact Stars. Nuclear Physics, Particle Physics and General
  Relativity.}}
\newblock Compact Stars.~ Nuclear Physics, Particle Physics and General
  Relativity, Approx.~390 pp.~90 figs..~Springer-Verlag New York.~ Also
  Astronomy and Astrophysics Library, 1996.

\bibitem{pie04}
J.~{Piekarewicz}.
\newblock {Unmasking the nuclear matter equation of state}.
\newblock {\em Physics Review C}, 69(4):041301--+, April 2004.

\bibitem{web07}
F.~{Weber}, R.~{Negreiros}, P.~{Rosenfield}, and M.~{Stejner}.
\newblock {Pulsars as astrophysical laboratories for nuclear and particle
  physics}.
\newblock {\em Progress in Particle and Nuclear Physics}, 59:94--113, July
  2007.

\bibitem{lat07}
J.~M. {Lattimer} and M.~{Prakash}.
\newblock {Neutron star observations: Prognosis for equation of state
  constraints}.
\newblock {\em Physics Reports}, 442:109--165, April 2007.

\bibitem{kla07}
T.~{Kl{\"a}hn}, D.~{Blaschke}, F.~{Sandin}, C.~{Fuchs}, A.~{Faessler},
  H.~{Grigorian}, G.~{R{\"o}pke}, and J.~{Tr{\"u}mper}.
\newblock {Modern compact star observations and the quark matter equation of
  state}.
\newblock {\em Physics Letters B}, 654:170--176, October 2007.

\bibitem{har06}
C.~{Hartnack}, H.~{Oeschler}, and J.~{Aichelin}.
\newblock {Recent astrophysical and accelerator-based results on the hadronic
  equation of state}.
\newblock {\em Journal of Physics G Nuclear Physics}, 32:231--+, December 2006.

\bibitem{for07}
A.~{F{\"o}rster}, F.~{Uhlig}, I.~{B{\"o}ttcher}, D.~{Brill}, M.~{de{\c
  B}owski}, F.~{Dohrmann}, E.~{Grosse}, P.~{Koczo{\'n}}, B.~{Kohlmeyer},
  S.~{Lang}, F.~{Laue}, M.~{Mang}, M.~{Menzel}, C.~{M{\"u}ntz}, L.~{Naumann},
  H.~{Oeschler}, M.~{P{\l}osko{\'n}}, W.~{Scheinast}, A.~{Schmah}, T.~J.
  {Schuck}, E.~{Schwab}, P.~{Senger}, Y.~{Shin}, J.~{Speer}, H.~{Str{\"o}bele},
  C.~{Sturm}, G.~{Sur{\'o}wka}, A.~{Wagner}, and W.~{Walu{\'s}}.
\newblock {Production of $K^{+}$ and of $K^{-}$ mesons in heavy-ion collisions
  from 0.6A to 2.0A GeV incident energy}.
\newblock {\em Physics Review C}, 75(2):024906--+, February 2007.

\bibitem{ada07}
A.~{Adare}, S.~{Afanasiev}, C.~{Aidala}, N.~N. {Ajitanand}, Y.~{Akiba},
  H.~{Al-Bataineh}, J.~{Alexander}, A.~{Al-Jamel}, K.~{Aoki}, (...), and
  L.~{Zolin}.
\newblock {Energy Loss and Flow of Heavy Quarks in Au+Au Collisions at
  $s_{NN}=200 {\rm GeV}$}.
\newblock {\em Physical Review Letters}, 98(17):172301--+, April 2007.

\bibitem{mat07}
D.~{Mateos}, R.~C. {Myers}, and R.~M. {Thomson}.
\newblock {Holographic Viscosity of Fundamental Matter}.
\newblock {\em Physical Review Letters}, 98(10):101601--+, March 2007.

\bibitem{mcc90}
P.~M. {McCulloch}, P.~A. {Hamilton}, D.~{McConnell}, and E.~A. {King}.
\newblock {The VELA glitch of Christmas 1988}.
\newblock {\em Nature}, 346:822--824, August 1990.

\bibitem{won01}
T.~{Wong}, D.~C. {Backer}, and A.~G. {Lyne}.
\newblock {Observations of a Series of Six Recent Glitches in the Crab Pulsar}.
\newblock {\em Astrophysical Journal}, 548:447--459, February 2001.

\bibitem{wal69}
G.~{Walin}.
\newblock {Some aspects of time-dependent motion of a stratified rotating
  fluid}.
\newblock {\em Journal of Fluid Mechanics}, 36:289--307, 1969.

\bibitem{bil00}
L.~{Bildsten} and G.~{Ushomirsky}.
\newblock {Viscous Boundary-Layer Damping of R-Modes in Neutron Stars}.
\newblock {\em Astrophysical Journall}, 529:L33--L36, January 2000.

\bibitem{kha65}
I.~M. {Khalatnikov} and D.~M. {Chernikova}.
\newblock {Dispersion of Sound in Superfluid Helium}.
\newblock {\em Soviet Journal of Experimental and Theoretical Physics Letters},
  2:351--+, December 1965.

\bibitem{rei93}
A.~{Reisenegger}.
\newblock {The spin-up problem in He II}.
\newblock {\em J.~Low Temp.~Phys.}, 92:77--106, 1993.

\bibitem{bar00}
C.~F. {Barenghi}.
\newblock {Superfluid Couette flow}.
\newblock In C.~{Egbers} and G.~{Pfister}, editors, {\em Physics of Rotating
  Fluids}, volume 549 of {\em Lecture Notes in Physics, Berlin Springer
  Verlag}, pages 379--+, 2000.

\bibitem{men01}
G.~{Mendell}.
\newblock {Magnetic effects on the viscous boundary layer damping of the
  r-modes in neutron stars}.
\newblock {\em Physics Review D}, 64(4):044009--+, August 2001.

\bibitem{alp96}
M.~A. {Alpar}, H.~F. {Chau}, K.~S. {Cheng}, and D.~{Pines}.
\newblock {Postglitch Relaxation of the Crab Pulsar after Its First Four Major
  Glitches: The Combined Effects of Crust Cracking, Formation of Vortex
  Depletion Region and Vortex Creep}.
\newblock {\em Astrophysical Journal}, 459:706--+, March 1996.

\bibitem{gre68}
H.~P. {Greenspan}, editor.
\newblock {\em {The Theory of Rotating Fluids}}, 1968.

\bibitem{mel07b}
A.~{Melatos} and C.~{Peralta}.
\newblock {Superfluid Turbulence and Pulsar Glitch Statistics}.
\newblock {\em Astrophysical Journall}, 662:L99--L102, June 2007.

\bibitem{ped83}
J.~{Pedlosky} and J.~T.~F. {Zimmerman}.
\newblock {Book-Review - Geophysical Fluid Dynamics}.
\newblock {\em Space Science Reviews}, 36:425--+, December 1983.

\bibitem{men98}
G.~{Mendell}.
\newblock {Magnetohydrodynamics in superconducting-superfluid neutron stars}.
\newblock {\em Monthly Notices of the Royal Astronomical Society},
  296:903--912, June 1998.

\bibitem{ped67}
J.~{Pedlosky}.
\newblock {The spin up of a stratified fluid}.
\newblock {\em Journal of Fluid Mechanics}, 28:463--479, 1967.

\bibitem{mou07}
J.~{Mound} and B.~{Buffett}.
\newblock {Viscosity of the Earth's fluid core and torsional oscillations}.
\newblock {\em Journal of Geophysical Research (Solid Earth)}, 112:5402--+, May
  2007.

\bibitem{and01b}
N.~{Andersson} and G.~L. {Comer}.
\newblock {On the dynamics of superfluid neutron star cores}.
\newblock {\em Monthly Notices of the Royal Astronomical Society},
  328:1129--1143, December 2001.

\bibitem{gre63}
H.~P. {Greenspan} and L.~N. {Howard}.
\newblock {On a time-dependent motion of a rotating fluid}.
\newblock {\em Journal of Fluid Mechanics}, 17:385--404, 1963.

\bibitem{hil77}
R.~N. {Hills} and P.~H. {Roberts}.
\newblock {Superfluid mechanics for a high density of vortex lines}.
\newblock {\em Archive for Rational Mechanics and Analysis}, 66:43--71, March
  1977.

\bibitem{war08}
L.~{Warszawski} and A.~{Melatos}.
\newblock {}.
\newblock {\em In preparation}, 2008.

\bibitem{ben08}
{Van Eysden}~C.~A. {Bennett}, M. and A.~{Melatos}.
\newblock {}.
\newblock {\em In preparation}, 2008.

\bibitem{stu01}
C.~{Sturm}, I.~{B{\"o}ttcher}, M.~{D{\c e}bowski}, A.~{F{\"o}rster},
  E.~{Grosse}, P.~{Koczo{\'n}}, B.~{Kohlmeyer}, F.~{Laue}, M.~{Mang},
  L.~{Naumann}, H.~{Oeschler}, F.~{P{\"u}hlhofer}, E.~{Schwab}, P.~{Senger},
  Y.~{Shin}, J.~{Speer}, H.~{Str{\"o}bele}, G.~{Sur{\'o}wka}, F.~{Uhlig},
  A.~{Wagner}, and W.~{Walu{\'s}}.
\newblock {Evidence for a Soft Nuclear Equation-of-State from Kaon Production
  in Heavy-Ion Collisions}.
\newblock {\em Physical Review Letters}, 86:39--42, January 2001.

\bibitem{fuc01}
C.~{Fuchs}, A.~{Faessler}, E.~{Zabrodin}, and Y.-M. {Zheng}.
\newblock {Probing the Nuclear Equation of State by $K^{+}$ Production in
  Heavy-Ion Collisions}.
\newblock {\em Physical Review Letters}, 86:1974--1977, March 2001.

\bibitem{adl03}
S.~S. {Adler}, S.~{Afanasiev}, C.~{Aidala}, N.~N. {Ajitanand}, Y.~{Akiba},
  J.~{Alexander}, R.~{Amirikas}, L.~{Aphecetche}, S.~H. {Aronson}, (...), and
  L.~{Zolin}.
\newblock {Elliptic Flow of Identified Hadrons in Au+Au Collisions at
  $s_{NN}=200 {\rm GeV}$}.
\newblock {\em Physical Review Letters}, 91(18):182301--+, October 2003.

\bibitem{kov05}
P.~K. {Kovtun}, D.~T. {Son}, and A.~O. {Starinets}.
\newblock {Viscosity in Strongly Interacting Quantum Field Theories from Black
  Hole Physics}.
\newblock {\em Physical Review Letters}, 94(11):111601--+, March 2005.

\bibitem{cut87}
C.~{Cutler} and L.~{Lindblom}.
\newblock {The effect of viscosity on neutron star oscillations}.
\newblock {\em Astrophysical Journal}, 314:234--241, March 1987.

\bibitem{abb04}
B.~{Abbott}, R.~{Abbott}, R.~{Adhikari}, A.~{Ageev}, B.~{Allen}, R.~{Amin},
  S.~B. {Anderson}, W.~G. {Anderson}, M.~{Araya}, (...), and J.~{Zweizig}.
\newblock {Analysis of first LIGO science data for stochastic gravitational
  waves}.
\newblock {\em Physics Review D}, 69(12):122004--+, June 2004.

\end{thebibliography}

\end{document}